\documentclass[12pt,final]{iopart}
\pdfoutput=1

\usepackage[utf8]{inputenc}

\usepackage{graphicx}
\usepackage[center]{subfigure}

\newcommand{\deriv}[2]{\ensuremath{\frac{\partial #1}{\partial #2}}}

\newcommand{\apar}{\ensuremath{A_{||}}}

\newcommand{\Vec}[1]{\ensuremath{\mathbf{#1}}}
\newcommand{\bvec}{\Vec{b}}
\newcommand{\kvec}{\Vec{\kappa}}
\newcommand{\bdotGrad}{\left(\bvec\cdot\nabla\right)}
\newcommand{\bxk}{\bvec_0\times\kvec_0\cdot\nabla}

\newcommand{\Jpar}{J_{||}}

\newcommand{\delp}{\nabla_\perp^2}

\begin{document}

\title[BOUT++ verification]{Verification of BOUT++ by the Method of Manufactured Solutions}
\author{B.D.Dudson$^1$, J.Madsen$^2$, J.Omotani$^3$, P.Hill$^{4,1}$, L.Easy$^{1,3}$, M.L\o{}iten$^2$}
\address{$^1$ York Plasma Institute, Department of Physics, University of York, Heslington, York YO10 5DD, UK \ead{benjamin.dudson@york.ac.uk}}
\address{$^2$ Department of Physics, Technical University of Denmark, DK-2800 Kgs. Lyngby, Denmark}
\address{$^3$ CCFE, Culham Science Centre, Abingdon, OX14 3DB, UK}
\address{$^4$ IRFN, CEA Cadarache, 13108 St Paul lez Durance, France}

\begin{abstract}

BOUT++ is a software package designed for solving plasma fluid models. It has been used to simulate a wide range of plasma
phenomena ranging from linear stability analysis to 3D plasma turbulence, and is capable of simulating a wide range of
drift-reduced plasma fluid and gyro-fluid models. 
A verification exercise has been performed as part of a EUROfusion Enabling Research project, to rigorously
test the correctness of the algorithms implemented in BOUT++, by testing order-of-accuracy convergence rates
using the Method of Manufactured Solutions (MMS). We present tests of individual components including time-integration
and advection schemes, non-orthogonal coordinate systems and the shifted metric procedure which is used to handle highly sheared grids.
The Flux Coordinate Independent (FCI) approach to differencing along magnetic field-lines has been implemented
in BOUT++, and is here verified using the MMS in a sheared slab configuration. Finally we show tests
of three complete models: 2-field Hasegawa-Wakatani, 3-field reduced MHD in 3D toroidal coordinates, and 5-field reduced MHD
in slab geometry.
\end{abstract}


\section{Introduction}
\label{sec:intro}

The BOUT++ code~\cite{Dudson2009,dudson2014} is an open source toolkit for the simulation of plasma models. Its applications include the study of plasma transients including Edge Localised Modes and filament / blob transport, and turbulence in magnetised plasma devices. Here we present a rigorous code verification exercise~\cite{roache1998,oberkampf2010} of the BOUT++ core algorithms and numerical methods, using the Method of Manufactured Solutions (MMS)~\cite{roache1998,salari2000}.
Code verification is a process of checking that the chosen set of partial differential equations is solved correctly
and consistently, and is a purely mathematical exercise. Code verification is not concerned with verifying that the chosen numerical methods are appropriate for the chosen set of equations. Code verification is also not concerned with testing the ability of a given model to explain experimental observations. This testing is dealt with in the subsequent validation process.
Code verification tests typically rely on a known solution against which to check the result (the Method of Exact Solutions). In relatively simple geometries (e.g. slabs or cylinders) and equations (usually linearised) an analytical solution can sometimes be found, and this kind of test is used to verify BOUT~\cite{umansky-2008-tests} and BOUT++~\cite{Dudson2009} as part of a test suite, run regularly to reduce the chances of errors being introduced. The requirement that there be an analytical solution restricts the usefulness of the tests, as the code cannot be verified for realistic geometries and problems of interest, where no such exact solution exists. 

The Method of Manufactured Solutions (MMS)~\cite{roache1998,salari2000} provides a method by which a simulation code can be verified in general situations, even where analytic solutions cannot be found. This is done by imposing a known ``manufactured'' solution, and adding sources to the equations such that the manufactured solution is an exact solution to the modified set of equations. 
The manufactured solution and therefore also the source are composed of primitive analytical functions $\sin$, $\cos$, $\tanh$ etc. 
which can be evaluated with a very high accuracy, typically double floating point precision. The difference between the 
numerically calculated solution and the ``exact'' manufactured solution provides the numerical error. The scaling of the
numerical error with the numerical spatial resolution is known {\it a priori}, and hence any deviation from the theoretical
scaling must be due to code inconsistencies or errors. 
The MMS is a very general technique, which has been used to verify a wide range of engineering codes, particularly in the fluid dynamics community~\cite{roy2004}. MMS has been applied to components of plasma simulation codes such as the European Transport Solver~\cite{kalupin2008}, gyrokinetic simulations~\cite{chang2009}, and has recently been applied to the GBS turbulence code~\cite{riva2014} and tokamak edge simulations~\cite{michoski2014}. 

As in~\cite{riva2014}, here we focus on order-of-accuracy tests as they provide the most rigorous test of numerical implementation~\cite{oberkampf2010}.
In section~\ref{sec:testing} we describe in more detail the MMS procedure, and the changes made to BOUT++ to facilitate its routine use. BOUT++ simulations typically employ non-orthogonal curvilinear coordinate systems, which are described in section~\ref{sec:coordinates} along with the method used to perform tests in these coordinates. Individual components of BOUT++ are first tested independently, including time integration schemes in section~\ref{sec:timeint}, advection schemes in section~\ref{sec:advection}, and operators for wave and diffusion equations along magnetic fields in section~\ref{sec:parallel}. Coordinate systems are then tested in section~\ref{sec:testcoords}. In section~\ref{sec:models} complete models are tested, in which these components are combined: The 2-field Hasegawa-Wakatani model of drift-wave turbulence in section~\ref{sec:hw}; a 3-field reduced Magnetohydrodynamics (MHD) model in section~\ref{sec:3field}; and a 5-field reduced MHD model similar to that in \cite{riva2014} is tested in section~\ref{sec:5field}.

All source code, input files, and scripts needed to produce the figures
and results in this paper are publicly available as part of the BOUT++ 
development repository at \texttt{https://github.com/boutproject/BOUT-dev}, 
revision 83c1f53. Due to automation of the testing procedure (section~\ref{sec:testing}), most results and figures in this paper
can be reproduced by running a single Python script. The location of these scripts will be specified relative to the root of the git repository.

\section{Testing framework}
\label{sec:testing}

The BOUT++ code is not limited to a single set of equations, but has been developed
to allow an arbitrary number of evolving fields, and input of
custom evolution equations in a form close to mathematical notation (see~\cite{Dudson2009,dudson2014} for details). 
This flexibility presents a challenge for verification, due to the large number of possible
combinations of operators and settings such as boundary conditions, which could be employed. Fortunately, as pointed out in~\cite{salari2000}, only mutually exclusive settings and operators need be independently tested, not all possible combinations of options. 
This still requires a relatively large number of tests to adequately cover the code components, and to verify each model. The process of MMS testing has therefore been automated as far as possible, by enabling all aspects of the test to be specified in an input text file. This allows the same code to be tested with different inputs, and new tests to be created more easily. Here we briefly outline the MMS procedure, before describing the mechanisms implemented in BOUT++ to carry out MMS testing.

Time integration codes such as BOUT++ evolve a set of nonlinear equations
for quantities $\underline{f}$, e.g. for a two field model evolving particle density and temperature $\underline{f}=\left\{n,T\right\}$. The system of equations is solved using the Method of Lines, and  can be written in a general form as:
\begin{equation}
\frac{\partial\underline{f}}{\partial t} = F\left(\underline{f}\right)
\end{equation}
where $F\left(\cdot\right)$ is a nonlinear operator which contains discretised differential operators in the spatial dimensions. In order to test the correctness of the numerical implementation, a time-dependent function $\underline{f}^M\left(t\right)$ is chosen (manufactured) using a combination of primitive mathematical functions which can be evaluated to machine precision. 
Manufactured solutions should be chosen so that they exercise all parts of the code, so should be varying in time and all spatial dimensions. Ideally the magnitude of the terms in the equations solved should be comparable, so that the error in one does not dominate over the others. Since derivatives of the solution will be taken numerically, the solution should also be smooth. 
Where the domain is periodic, such as toroidal angle in tokamak simulations, the manufactured solutions must also be periodic in those directions. A detailed discussion of selection criteria for manufactured solutions can be found in~\cite{salari2000}.

The manufactured function $\underline{f}^M$ is now inserted into the function $F\left(\cdot\right)$ and $\frac{\partial\underline{f}^M}{\partial t}$ to calculate a source function $S$ analytically:
\begin{equation}
S\left(t\right) = \frac{\partial\underline{f}^M}{\partial t} - F\left(\underline{f}^M\right)
\label{eq:mms_source}
\end{equation}
Here the symbolic packages Mathematica and the Sympy library~\cite{sympy} were used to calculate source
functions. Both can generate representations of the resulting expressions which
can be copied directly into source code or text input files. For large sets of equations such as those in section~\ref{sec:5field} this is essential in order to avoid introducing errors.

The system of equations to be solved numerically is now modified to:
\begin{equation}
\frac{\partial\underline{f}}{\partial t} = F\left(\underline{f}\right) + S\left(t\right)
\label{eq:mmstime}
\end{equation}
so that the function $\underline{f}^M$ is an exact (manufactured) solution of equation~\ref{eq:mmstime}. Since $S$ has been calculated analytically, it can be evaluated to within machine precision at any desired time, and passed to the time integration routines. At the start of the simulation $t=0$ the state is set to the manufactured solution $\underline{f} = \underline{f}^M\left(t=0\right)$. The simulation time is then advanced to some later time $t = \Delta t$, at which point the numerical solution $\underline{f}$ is compared to the manufactured solution $\underline{f}^M\left(t=\Delta t\right)$. The norm of the difference between the numerical solution and the manufactured solution $\epsilon = \left|\left|\underline{f} - \underline{f}^M\right|\right|$ at $t = \Delta t$ then gives a measure of the error in the numerical solution, which should converge towards zero as the spatial and temporal mesh is refined. Note that in order to obtain convergence in the solution of a time-dependent Partial Differential Equation (PDE), both the spatial and temporal mesh (time step) must be refined~\cite{leveque2007}. In general separating the spatial and temporal convergence is non-trivial, but in section~\ref{sec:secondderiv} we use a slightly different procedure than outlined above, to verify spatial convergence and boundary conditions independently of temporal convergence.

Boundary conditions must also be modified for testing with the MMS. A Dirichlet boundary condition on a quantity $n$ (e.g. particle density), for example, must be modified to set the solution equal to the time-varying manufactured solution $n^M$ on the boundary:
\begin{equation}
n\left(\mathrm{boundary}\right) = n^M\left(\mathrm{boundary}, t\right)
\end{equation}
Similarly for Neumann boundary conditions: 
\begin{equation}
\frac{\partial n}{\partial x}\left(\mathrm{boundary}\right) = \frac{\partial n^M}{\partial x}\left(\mathrm{boundary}, t\right)
\end{equation}
More complex boundary conditions such as sheaths, which couple multiple fields together, can be treated by adding a source function as for the time integration equation~\ref{eq:mmstime}.
The boundary conditions applied to all fields now become time-dependent, and must be evaluated from an analytic expression at arbitrary points in time.

In order to test a numerical model using the Method of Manufactured Solutions, three analytic function inputs are therefore required for each evolving field (e.g. density $n$, temperature $T$, ...):
\begin{enumerate}
\item A manufactured solution
\item A source function calculated from equation~\ref{eq:mms_source} using a symbolic package like SymPy
\item Analytic expressions for boundary conditions
\end{enumerate}
As described in~\cite{dudson2014}, BOUT++ contains an expression parser which evaluates analytic expressions in input files. This was added as a convenient means to specify initial conditions, but has been extended and adapted for use in MMS testing. Once MMS testing is enabled by setting a flag in the input, BOUT++ reads a manufactured solution from the input for each evolving variable, using it to initialise the variable and to calculate an error at each output time; a source function is read and used to modify the time derivatives which are passed to the time-integration code; and expressions for boundary conditions are evaluated at the required times. All of this machinery is independent of the specific model, and in most cases does not require any modification of the problem-specific code\footnote{The only code changes required for MMS testing are Laplacian inversions, which currently require some modifications to their calls in order to insert additional source functions}. The form of the analytic expressions is of course problem specific, but once calculated, a BOUT++ executable can be tested using MMS and then used to perform physics simulations without recompiling, only changing the input file. This automation of the testing process aims to lower the barriers to routine testing of BOUT++ simulation models using the Method of Manufactured Solutions. 

\section{Coordinate systems}
\label{sec:coordinates}

In strongly magnetized plasmas the characteristic gradient length scales parallel to the magnetic field are often much longer
than the perpendicular length scales. This scale separation is often exploited in numerical simulation to reduce
the computational cost by using a coarser discretisation in the direction parallel to the magnetic field. A widely used approach
is to express the model equations in magnetic field-aligned, curvilinear coordinates. In most previous BOUT++
simulations~\cite{Dudson2009} we have used the so-called ballooning coordinates. Starting from orthogonal toroidal flux coordinates~\cite{haeseler-1} $\left(\psi, \theta, \zeta\right)$ with radial flux-surface label $\psi$, poloidal angle $\theta$, and toroidal angle $\zeta$, the coordinates are transformed to field-aligned ballooning coordinates $\left(x,y,z\right)$~\cite{xu-2008}
\begin{equation}
x = \psi \qquad y = \theta \qquad z = \zeta - \int_{\theta_0}^\theta \nu d\theta \qquad \nu = \frac{B_\zeta r}{B_\theta R}
\label{eq:ballooning}
\end{equation}
where $B_\zeta$ and $B_\theta$ are the toroidal and poloidal magnetic field components, $r$ is the minor radius, $R$ is the major radius, and $\nu$ is the local magnetic field-line pitch. Moving along $y$ at fixed $x$ and $z$ follows the path of a field-line in both $\theta$ and $\zeta$. The covariant basis vector (the vector between grid-points) is~\cite{Dudson2009}:
\begin{eqnarray}
\mathbf{e}_x &=& \frac{1}{RB_\theta}\mathbf{\hat{e}}_\psi + IR\mathbf{\hat{e}}_\zeta \label{eq:cov_basis} \\
\mathbf{e}_y &=& \frac{h_\theta}{B_\theta}\mathbf{B} \nonumber \\
\mathbf{e}_z &=& R\mathbf{\hat{e}}_\zeta \nonumber
\end{eqnarray}
where $\mathbf{\hat{e}}$ are the unit vectors in the original orthogonal toroidal $\left(\psi, \theta, \zeta\right)$
coordinate system, and $I = \int\frac{\partial \nu}{\partial \psi}d\theta$ is the integrated shear. The magnetic field is given by $\mathbf{B} = \nabla z \times \nabla x$, and so the derivative along the magnetic field reduces to a simple partial derivative $\mathbf{B}\cdot\nabla = \nabla z\times\nabla x \cdot \nabla y \partial_y$. Since fluctuations typically have long wavelengths along field-lines, a lower resolution can be used in this parallel $y$ coordinate, with a corresponding reduction in computational resources, both run time and memory.

In order to reduce the deformation of the coordinates caused by
magnetic shear $I$ (see $\mathbf{e}_x$ in equation~\ref{eq:cov_basis}), a shifted metric method~\cite{dimits-1993,scott01}
is usually used, a discussion of which can be found in~\cite{Dudson2009}
and more recently in~\cite{hariri2013}. At each $y = $~const plane, a local coordinate system is defined in which $x$ and $z$ are orthogonal. Mapping
between these local coordinates and the global field-aligned coordinates can be done using Fast Fourier Transforms (FFTs) in the toroidal ($\zeta$, $z$) 
direction. As implemented in BOUT++, this procedure has no effect on differencing in the parallel direction, but differencing in $x$ is modified
by shifting quantities in $z$ using FFTs before calculating finite differences. 

A toroidal coordinate system for MMS testing is generated by first specifying
the path of magnetic field lines in poloidal and toroidal angle.
The poloidal magnetic field can then be calculated by differentiation, ensuring
that the resulting analytic metric tensor components have relatively
compact closed forms. The formula used here for the toroidal angle $\zeta$ as a function of the radial (flux) coordinate $\psi$ and poloidal angle $\theta$ is:
\begin{equation}
\zeta = q\left(\psi\right)\left[ \theta + \epsilon \sin\theta \right]
\label{eq:fieldline}
\end{equation}
where $q\left(\psi\right)$ is the safety factor, which is taken to be a parabolic function of $\psi$ varying between $2$ and $3$ in sections~\ref{sec:testcoords} and~\ref{sec:3field}. $\epsilon = r / R_0$ is the inverse aspect ratio, here taken to be $\epsilon = 0.1$. 
From this, the field line pitch is calculated as
\begin{equation}
\nu = q\left(\psi\right)\left[ 1 + \epsilon \cos\theta \right]
\label{eq:fieldpitch}
\end{equation}
A fixed value of the poloidal current function $f = B_\zeta R$ and minor radius $r = \epsilon R_0$ 
is used, and the major radius of a field line varies as $R = R_0 + r\cos\theta$. Equation~\ref{eq:fieldpitch} is then rearranged to give an expression for the poloidal field. The integrated shear is calculated from the differential of the field-line toroidal angle $\zeta$ with respect to $\psi$:
\begin{equation}
I = \int_{\theta_0}^\theta\frac{\partial\nu}{\partial\psi}d\theta = \frac{\partial q\left(\psi\right)}{\partial \psi} \left[ 1 + \epsilon \cos\theta \right]
\label{eq:sinty}
\end{equation}

The resulting covariant and contravariant metric tensors have the same non-zero pattern as in simulations of real devices, and elements of the covariant metric tensor vary in both radial and poloidal coordinates~\cite{xu-2008}.
Differencing operators parallel and perpendicular to the magnetic field are tested in this coordinate system in section~\ref{sec:testcoords},
and a 3-field electromagnetic reduced MHD model is verified in this coordinate system in section~\ref{sec:3field}.

\subsection{Flux Coordinate Independent scheme}

Recently a new approach to plasma turbulence simulations has been developed~\cite{hariri2013,stegmeir2014}, and work is ongoing to implement this scheme in several simulation codes. We have implemented this Flux Coordinate Independent (FCI) scheme in BOUT++, enabling the development of complex turbulence models in arbitrary magnetic geometry. By assuming that the poloidal plane equals the plane perpendicular to the magnetic field, complex non-orthogonal curvilinear field-aligned flux coordinates do not need to be used in the perpendicular direction, but can use simple geometries (e.g. Cartesian).
Here we verify that these numerical schemes have been implemented correctly
for a sheared slab geometry. Further development and verification in more complex geometries will be the subject of a future publication.

\begin{figure}[htbp!]
\centering
  \includegraphics[width=0.6\textwidth]{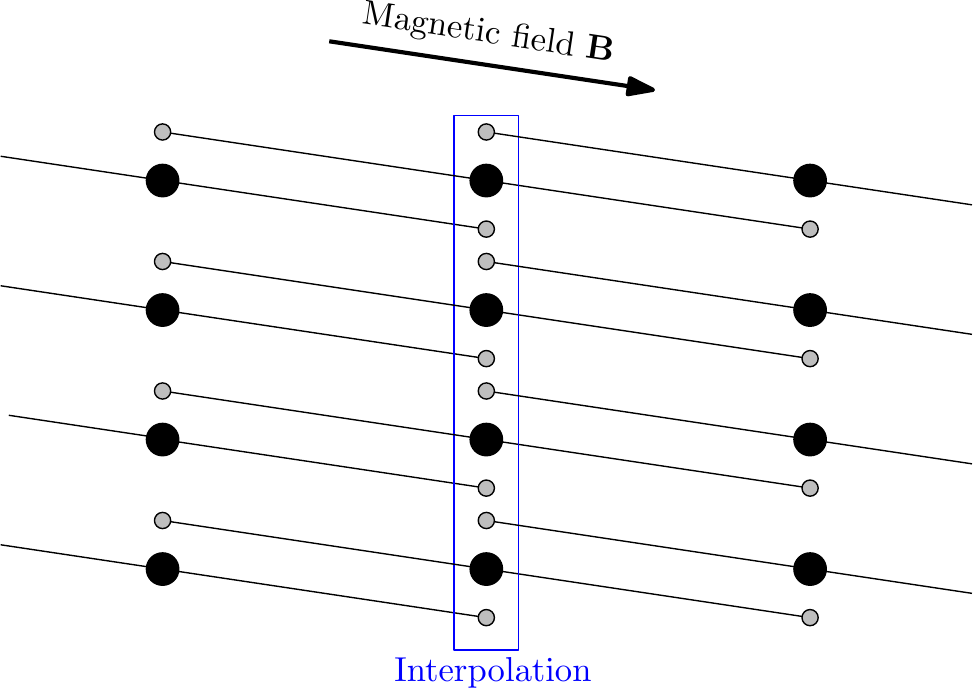}
\caption{Flux Coordinate Independent (FCI) scheme. To calculate the derivative along the magnetic field at grid cells (large solid circles), field-lines are followed in both directions to points on neighbouring planes of grid cells (small open circles). Values at these points are found by high-order interpolation using nearby points (blue box).}
\label{fig:fci_diagram}
\end{figure}
The Flux Coordinate Independent scheme, as implemented in BOUT++, employs 3$^{rd}$-order Hermite polynomial interpolation in the plane perpendicular to the magnetic field, and 2$^{nd}$-order central differencing along the magnetic field. The idea is illustrated in figure~\ref{fig:fci_diagram}: The grid is constructed to be dense in planes perpendicular to the magnetic field and sparse along the magnetic field, since from physical arguments we expect the solutions to vary slowly along magnetic field-lines ($k_{||} \ll k_\perp$). To calculate derivatives of a quantity $f$ along magnetic fields, the magnetic field is first followed
from each grid point onto neighbouring planes; values of $f$ on neighbouring planes are then interpolated onto these intersection locations. This gives
the value of $f$ at 3 points along the magnetic field (the starting grid point, and one point along the field in each direction), which
is sufficient to calculate second-order accurate first or second derivatives using central differencing. If higher order derivatives are required, then the magnetic field could be followed to calculate intersections with further planes. There are subtle issues with this scheme which will not be addressed here, and are left to future work: the treatment of boundary conditions where magnetic field-lines intersect material surfaces, and time-evolving magnetic fields
where the mapping of field-lines to neighbouring planes might need to be updated are two areas of interest. The efficiency of the scheme in terms of the computing time required for high-order interpolation is also important in determining the best overall scheme to employ, and is also left to future work.

\section{Results}

Since operators can be tested and verified independently (see~\cite{salari2000} and discussion in section~\ref{sec:testing}),
a suite of smaller tests is generally more useful than a test
which combines everything together. Whole models are tested in section~\ref{sec:models}, but require considerable computing resources to run, and if one of these fails then it is difficult to know where the error lies. Tests of individual components can run in minutes on a desktop, rather than hours on a supercomputer, and a test failure provides better guidance as to the location of the error. The difficulty is in the
large number of tests needed to ensure coverage: Here we verify the major components of BOUT++, including time integration schemes (section~\ref{sec:timeint}), advection operators (section~\ref{sec:advection}), central schemes for wave and diffusion equations (section~\ref{sec:parallel}), and the curvilinear coordinate system used
for tokamak simulations (section~\ref{sec:testcoords}). Other components, such as calculation of potential $\phi$ from vorticity, are verified
as part of full models (section~\ref{sec:models}), and development of individual tests for these components is a matter of ongoing work.

\subsection{Time integration}
\label{sec:timeint}

\begin{figure}[htbp!]
\centering
  \includegraphics[width=0.6\textwidth]{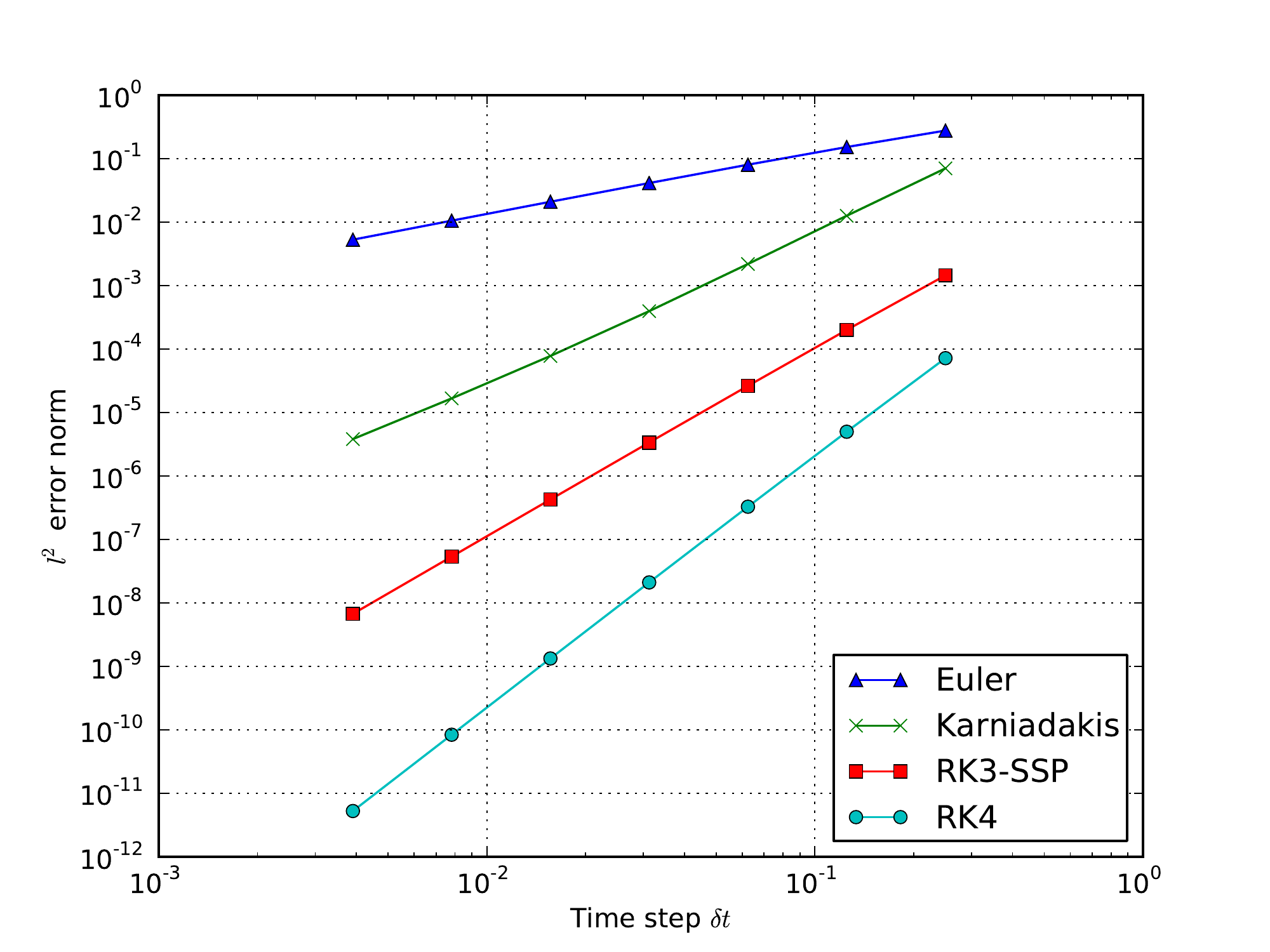}
\caption{Error norm for explicit time integration schemes. Measured convergence rates based on the two highest resolution cases are: 0.995 (Euler), 2.13 (Karniadakis), 3.00 (RK3-SSP), and 3.99 (RK4). Script:\texttt{examples/MMS/time/runtest}}
\label{fig:time_norm}
\end{figure}

Several explicit and implicit time integration schemes are implemented in BOUT++, 
allowing users to choose at run-time which scheme to use. Methods tested are the Euler, RK4~\cite{iserles2009}, a multi-step method derived by Karniadakis {\it et al}~\cite{karniadakis1991,scott2005-arxiv}, and a third-order Strong Stability Preserving Runge-Kutta method (RK3-SSP)~\cite{gottlieb2001}. Results obtained by integrating $\frac{\partial f}{\partial t} = f$ between
$t=0$ and $t=1$ are shown in figure~\ref{fig:time_norm}. Other functions such as $\frac{\partial f}{\partial t} = \cos(t)$ have also been tested, resulting in the same convergence rate.

The Euler, RK3-SSP and RK4 methods all reproduce their expected convergence rates (first, third, and fourth order in $\delta t$ respectively), and so can be considered verified. The Karniadakis scheme is expected to be third order accurate, but only second order convergence is observed. This is most likely due to the initialisation procedure of the multistep method: At each step the value of $f$ and its time derivative at two previous timesteps are required, and so to start the simulation these previous steps are constructed using Euler's method. This results in an $O\left(\delta t^2\right)$ error, reducing the overall convergence to second order in $\delta t$.

Time integration in BOUT++ simulations is typically done using implicit adaptive
Jacobian-Free Newton Krylov (JFNK) schemes, provided by either the SUite of Nonlinear and Differential/ALgebraic equation Solvers (SUNDIALS~\cite{hindmarsh2005}) or the Portable, Extensible Toolkit for Scientific Computation (PETSc~\cite{efficient,petsc-user-ref}). These use adaptive order and adaptive timesteps in order to achieve a user-specified tolerance, and so are difficult to validate using the MMS method.
Here we take as given that the time integration methods in these libraries are implemented correctly, and use SUNDIALS for time integration in the remainder of this paper with a relative tolerance of $10^{-8}$ and absolute tolerance of $10^{-12}$. These small tolerances are used so that the spatial discretisation error we are interested in dominates over the time integration error in the results which follow.

\subsection{Advection schemes}
\label{sec:advection}

A key component of drift-reduced plasma simulations are operators for
drifts across magnetic field-lines. These can be written in the form of
an advection equation, or as a Poisson bracket. For example the $E\times B$ drift
of a scalar quantity $f$ (e.g. density), due to an electrostatic potential $\phi$ is:
\begin{equation}
\frac{\partial f}{\partial t} = -\frac{1}{B}\mathbf{b}\times\nabla\phi\cdot \nabla f = -\left[\phi, f\right]
\end{equation}
Several schemes for calculating the Poisson bracket using both finite difference and finite volume discretizations are implemented in BOUT++. Some of these preserve the symmetries of the Poisson bracket (e.g. second order Arakawa~\cite{arakawa1960}), whilst others are designed to handle shocks and discontinuities robustly (e.g. WENO~\cite{jiang-1996,jiang-1997}). As with time integration schemes, users can switch between these methods at run-time. In order to test advection schemes, we simulate a single
scalar field $f$ advected by Poisson bracket using an imposed
potential $\phi$: 
\begin{equation}
\frac{\partial f}{\partial t} = -\left[\phi, f\right] - H \cdot \delta x^4 \nabla_\perp^4 f
\label{eq:advect}
\end{equation}
where $H$ is a hyper-diffusion constant, $\delta x$ is the mesh spacing, and the $\nabla_\perp^4$ operator is calculated using second-order central differences. The manufactured solutions were chosen to be:
\begin{eqnarray}
f &=& \cos\left(4\overline{x}^2 + \overline{z}\right) + \sin\left(t\right)\sin\left(3\overline{x} + 2\overline{z}\right)\label{eq:advectsoln} \\
\phi &=& \sin\left(6\overline{x}^2 - \overline{z}\right)
\end{eqnarray}
where the coordinates perpendicular to the magnetic field are normalised such that $0 \le \overline{x} \le 1$ and $0 \le \overline{z} \le 2\pi$. This solution varies smoothly in both $\overline{x}$ and $\overline{z}$, and in time. Note that the WENO scheme is a limiter scheme, which adapts its stencils depending on the local gradients, and this functionality is not properly tested here. Limiter and other adaptive schemes reduce accuracy in steep gradient regions in order to reduce or eliminate overshoot oscillations. This presents a challenge for MMS testing of convergence order, and as far as we are aware there is no accepted means of fully verifying these schemes using the MMS.

\begin{figure}[htbp!]
\centering
\subfigure[$l^2$ (RMS) error norm]{
  \label{fig:advection_norm_l2}
  \includegraphics[width=0.45\columnwidth]{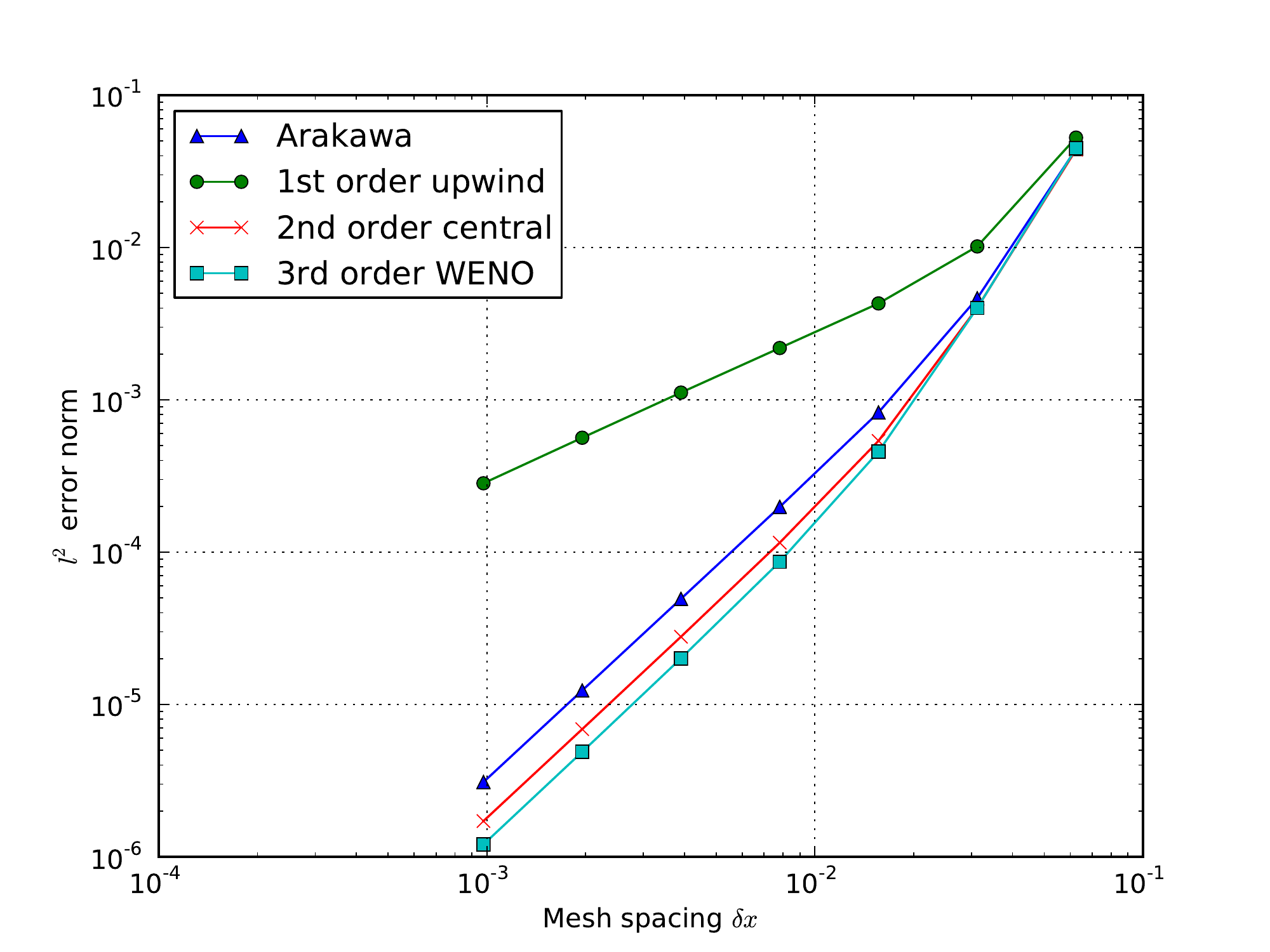}
}
\subfigure[$l^\infty$ (maximum) error norm]{
  \label{fig:advection_norm_linf}
  \includegraphics[width=0.45\columnwidth]{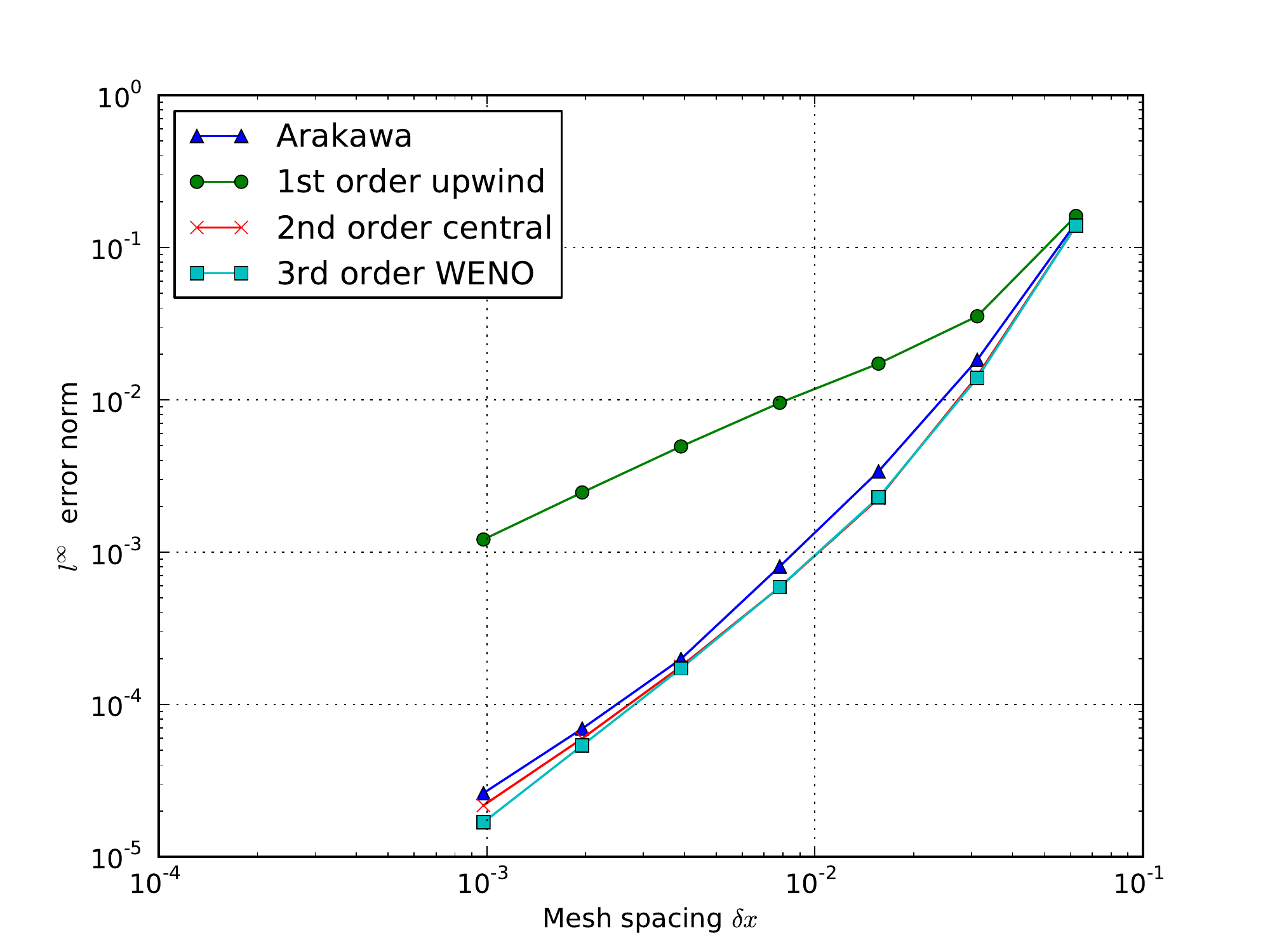}
}
\caption{MMS test of advection operators. Equation~\ref{eq:advect} is solved on a 2D domain with uniform grid spacing. The resolution varies from $16\times 16$ to $1024\times 1024$. Convergence rates for second-order Arakawa (1.998), 1$^{st}$-order upwind (0.993), 2$^{nd}$-order central differencing (2.005), and 3$^{rd}$-order WENO (2.019). All methods are limited to at best second-order in mesh spacing $\delta x$ due to the central differencing applied to $\phi$ and the boundary conditions. Script:\texttt{examples/MMS/advection/runtest}}
\label{fig:advection_norm}
\end{figure}

Advection schemes require some form of dissipation at the grid scale, in order to avoid numerical oscillations. In the upwind and WENO schemes this dissipation is provided by upwinding as part of the advection scheme itself, but central differencing schemes such as Arakawa have low dissipation, and require additional dissipation to stabilise the solution, either physically motivated or numerical. Since there is no other dissipation in this toy problem, a 4th-order hyper-diffusion term is added to equation~\ref{eq:advect}, with a coefficient $H$ which converges to zero at $\delta x^4$ for grid spacing $\delta x$. Without this dissipation term convergence is typically reduced to first order, and becomes dependent on the integration time due to the growth of numerical oscillations. When dissipation with $H=20$ is included, the results are shown in figure~\ref{fig:advection_norm}. 

Both global error and local error are found to converge at the expected rate in the asymptotic (small $\delta x$ regime, as measured by the $l^2$ (RMS) error in figure~\ref{fig:advection_norm_l2}, and the $l^\infty$ (maximum) error in figure~\ref{fig:advection_norm_linf} respectively. Apart from the first order upwind scheme, all schemes converge at second order in grid spacing $\delta x$: The WENO scheme is formally third order accurate in the bulk of the domain, but the advection velocity is calculated from $\phi$ using $2^{nd}$-order central differences, and boundary conditions are only second-order accurate, reducing the overall convergence rate to second order. The WENO scheme implementation cannot therefore be considered fully verified, and as noted above the verification of limiter schemes using MMS remains an outstanding problem, and so we leave this for further work.

\subsection{Schemes for wave equations}
\label{sec:parallel}

Along the magnetic field methods are implemented which model wave propagation, such
as sound and shear Alfv\'en waves, and diffusion processes such as heat conductivity.
Wave propagation operators often appear in the form of coupled first order equations:
\begin{equation}
\frac{\partial f}{\partial t} = \frac{\partial g}{\partial x} \qquad \frac{\partial g}{\partial t}  = \frac{\partial f}{\partial x}
\label{eq:parallel_wave}
\end{equation}
The manufactured solution was chosen to be
\begin{eqnarray*}
f = 0.9 + 0.9\overline{x} + 0.2\cos(10t)\sin(5\overline{x}^2) \nonumber \\
g = 0.9 + 0.7\overline{x} + 0.2\cos(7t)\sin(2\overline{x}^2)
\end{eqnarray*}
and the equations are solved using staggered 2$^{nd}$-order central differencing: Variable $g$ was shifted to the cell boundaries, whilst $f$ was cell centred. This arrangement requires different handling of boundary conditions to account for this shift. To test boundary conditions and handling of staggered variables, this test was performed in $x$ and then in $y$ (replacing $\overline{x}$ with $\overline{y}/2\pi$ in the above manufactured solutions).

\begin{figure}[htbp!]
\centering
\includegraphics[width=0.45\columnwidth]{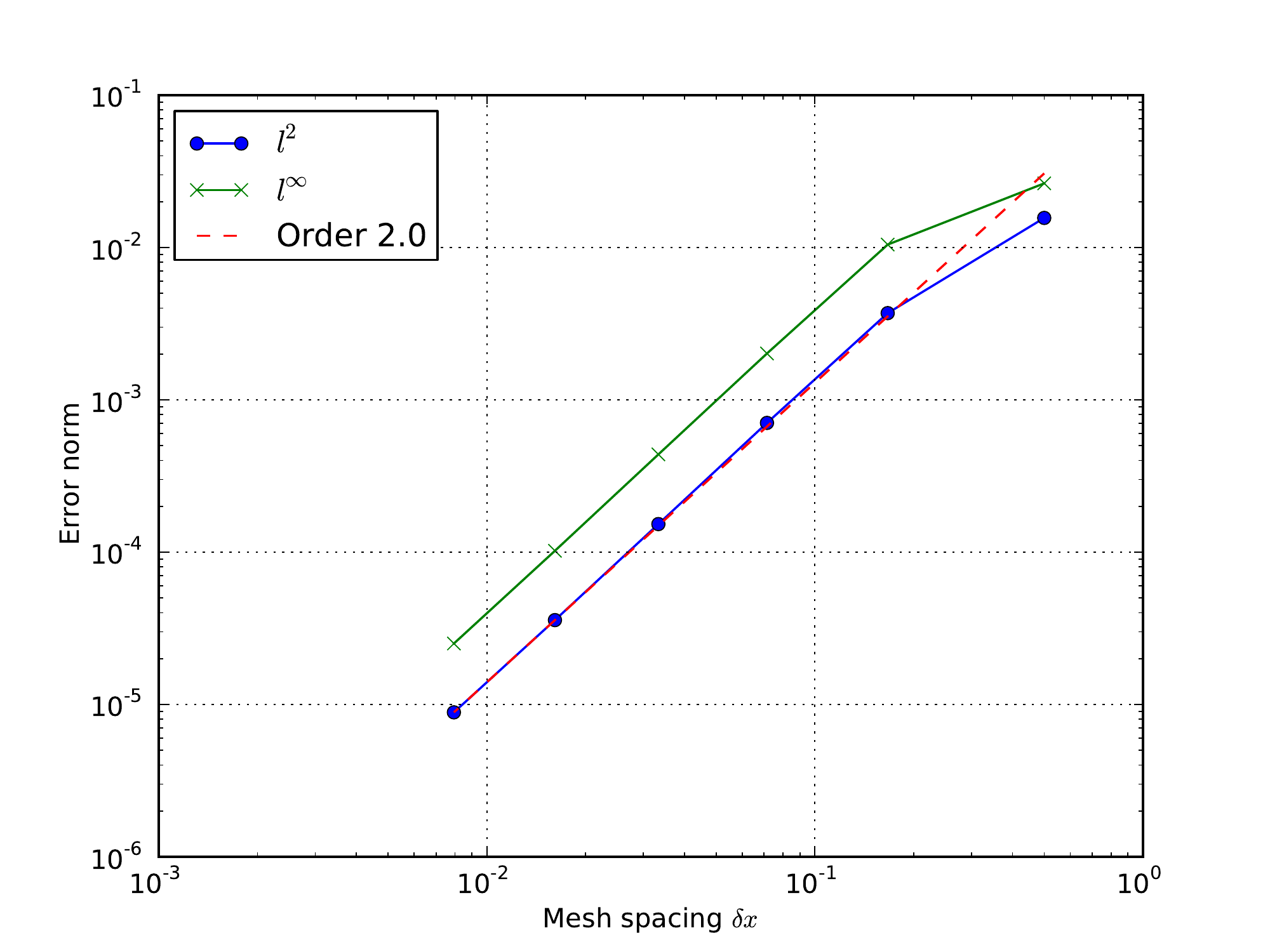}
\caption{Error norms for $f$ in wave equation~\ref{eq:parallel_wave} using $2^{nd}$-order central differencing. A convergence rate of $1.97$ is found, illustrated with a dashed line. Script:\texttt{examples/MMS/wave-1d/runtest}}
\label{fig:wave_norm}
\end{figure}

Results of a convergence test are shown in figure~\ref{fig:wave_norm},
which shows the $l^2$ (RMS) and $l^\infty$ (maximum) error norms for quantity
$f$ as a function of the mesh spacing $\delta x$. This shows convergence
at an order around $1.97$, as expected for this scheme. This test has been conducted with combinations of Dirichlet and Neumann boundary conditions, finding essentially the same result in all cases.

\subsection{Second derivative operators}
\label{sec:secondderiv}

In order to verify the second derivative (diffusive) operators and boundary conditions, a series of tests have been performed:
First we verify the spatial convergence rate towards a steady state (time-independent) manufactured solution; and then we verify using a
time-dependent manufactured solution. 

\subsubsection{Steady-state manufactured solution}

In order to verify spatial convergence for time-dependent systems of
equations, the approach taken in~\cite{salari2000} is to evolve the 
equations towards a steady-state solution. Here we use this approach
to verify boundary conditions and second-order operators by solving
the equation:
\begin{equation}
\frac{\partial f}{\partial t} = \frac{\partial^2 f}{\partial x^2} + S
\label{eq:diffusion1d}
\end{equation}
The manufactured solution is chosen to be
\begin{equation}
f^M = 0.9 + 0.9 x + 0.2\sin\left(5x^2\right)
\label{eq:diffusionsteady}
\end{equation}
in the range $0 \le x \le 1$ i.e. boundaries are at $x=0$ and $x = 1$. The source function is therefore:
\begin{equation}
S = 20x^2\sin\left(5x^2\right) - 2\cos\left(5x^2\right)
\end{equation}
In contrast to the time-dependent MMS tests presented in this paper,
for this steady-state problem we initialise the simulation at $t=0$ with
$f = 0$, and not the exact manufactured solution. This is suggested by~\cite{salari2000} since even though this increases the number of iterations to convergence, using the exact solution can hide coding mistakes. 
Equation~\ref{eq:diffusion1d} was then integrated in time to $t=10$ using
an absolute tolerance of $10^{-15}$ and relative tolerance of $10^{-7}$.
This is a sufficiently long time that $f$ reaches a steady state to within tolerances. 

Results are listed in table~\ref{tab:err}, showing $l^2$ and $l^\infty$ errors
and convergence rates. We first perform
the test with Dirichlet boundary conditions, then with mixed Dirichlet and Neumann conditions. In all cases 2$^{nd}$-order convergence is observed at high resolution.
\begin{table}[h]
\centering
\caption{Error norms and convergence rates for integration of equation~\ref{eq:diffusion} as a function of number of grid points $N$. Shown are cases with Dirichlet boundary conditions, then with one Dirichlet and one Neumann boundary (mixed).}
\label{tab:err}
\begin{tabular}{c c c c c c c c c}
  \hline
  \hline
   & \multicolumn{4}{c|}{Dirichlet} & \multicolumn{4}{c}{Mixed} \\
  \cline{2-5}\cline{6-9}
  $N$ & $l^2$ & Rate & $l^\infty$ & Rate & $l^2$ & Rate & $l^\infty$ & Rate \\
  8 & 2.624e-02 &  & 6.088e-02 & & 3.504e-02 &  & 6.317e-02 & \\
  16 & 4.332e-03 & 2.126 & 1.227e-02 & 1.890 & 5.514e-03 & 2.182 & 1.242e-02 & 1.919 \\
  32 & 9.224e-04 & 2.030 & 2.720e-03 & 1.978 & 1.165e-03 & 2.039 & 2.733e-03 & 1.986 \\
  64 & 2.149e-04 & 2.007 & 6.400e-04 & 1.993 & 2.712e-04 & 2.009 & 6.415e-04 & 1.997 \\
  128 & 5.199e-05 & 2.001 & 1.552e-04 & 1.997 & 6.554e-05 & 2.003 & 1.554e-04 & 1.999 \\
  256 & 1.271e-05 & 2.009 & 3.822e-05 & 1.999 & 1.607e-05 & 2.005 & 3.825e-05 & 2.000 \\
  512 & 3.395e-06 & 1.894 & 9.572e-06 & 1.986 & 4.000e-06 & 1.996 & 9.488e-06 & 2.000 \\
  \hline
  \hline
\end{tabular}
\end{table}

\subsubsection{Time-dependent manufactured solution}

\begin{figure}[htbp!]
  \centering
  \includegraphics[width=0.45\columnwidth]{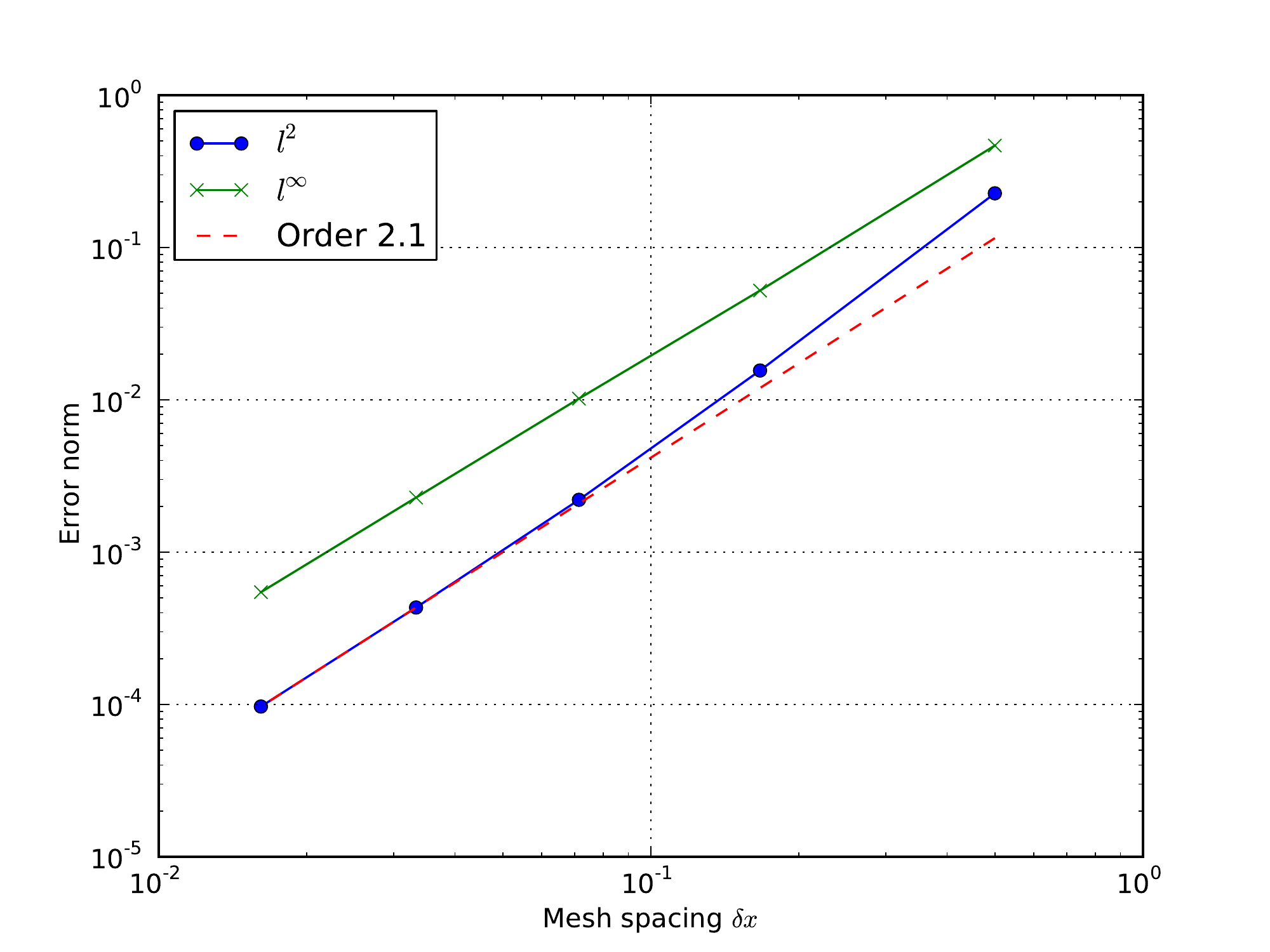}
  \caption{Error norms for diffusion equation~\ref{eq:diffusion} in 3D on a uniform grid as a function of grid spacing $\delta x$, showing convergence with an order of $2.06$. Script:\texttt{examples/MMS/diffusion2/runtest}}
  \label{fig:diffusion_norm}
\end{figure}

Diffusion equations in all three dimensions, separately and in combination, have been
verified, with convergence for one example shown in figure~\ref{fig:diffusion_norm}. The equation solved is
\begin{equation}
\frac{\partial f}{\partial t} = \nabla^2 f
\label{eq:diffusion}
\end{equation}
which is solved using 2$^{nd}$-order central differences on a uniform grid. 
In 3D the manufacutured solution used was
\begin{equation}
f = 0.9 + 0.9x + 0.2\cos\left(10t\right)\sin\left(5x^2 - 2z\right) + \cos\left(y\right)
\end{equation}
in the range $0 \le x \le 1$; $0 \le y \le 2\pi$ and $0 \le z \le 2\pi$.
Results for a uniform 3D grid are shown in figure~\ref{fig:diffusion_norm}, showing convergence at the expected order. 

These tests confirm that these simple operators and the Dirichlet and Neumann boundary conditions have been implemented correctly for uniform orthogonal grids. More complicated geometries are tested in the next section, but the advantage of these simple tests is that they run in under a minute on a desktop and so are now included in the standard BOUT++ test suite which is run routinely to check for errors.

\subsection{Coordinate systems}
\label{sec:testcoords}

The field-aligned coordinate system used for tokamak simulations has
been tested using the analytic input mesh described in section~\ref{sec:coordinates}.
The manufactured solution was
\begin{equation}
f = \cos\left(4\overline{x}^4 + \zeta - \theta\right) + \sin\left(t\right)\sin\left(3\overline{x} + 2\zeta - \theta\right)
\end{equation}
where $\overline{x} = \psi / \Delta \psi$ is a normalised radial coordinate with a range between $0$ and $1$. The safety factor was chosen to be $q = 2 + \overline{x}^2$, and inverse aspect ratio $\epsilon = 0.1$.
Following the procedure outlined in section~\ref{sec:coordinates}, this results in toroidal and poloidal magnetic field components:
\begin{eqnarray}
B_\zeta &=& \frac{1}{1 - 0.1\cos\left(\theta\right)} \\
B_\theta &=& \frac{0.1}{\left(\overline{x}^2+2\right)\left[1-0.1\cos\left(\theta\right)\right]^2\left(1 + 0.1\cos\left(\theta\right)\right)} \nonumber
\end{eqnarray}
and integrated shear
\begin{equation}
I = 1125x\left[\theta + 0.1\sin\left(\theta\right)\right]
\end{equation}

\begin{figure}[htbp!]
\centering
\subfigure[Ballooning coordinates. Convergence rates for Arakawa bracket (2.03); perpendicular diffusion operator (1.97); and parallel diffusion operator (1.99)]{
  \label{fig:ballooning}
  \includegraphics[width=0.45\columnwidth]{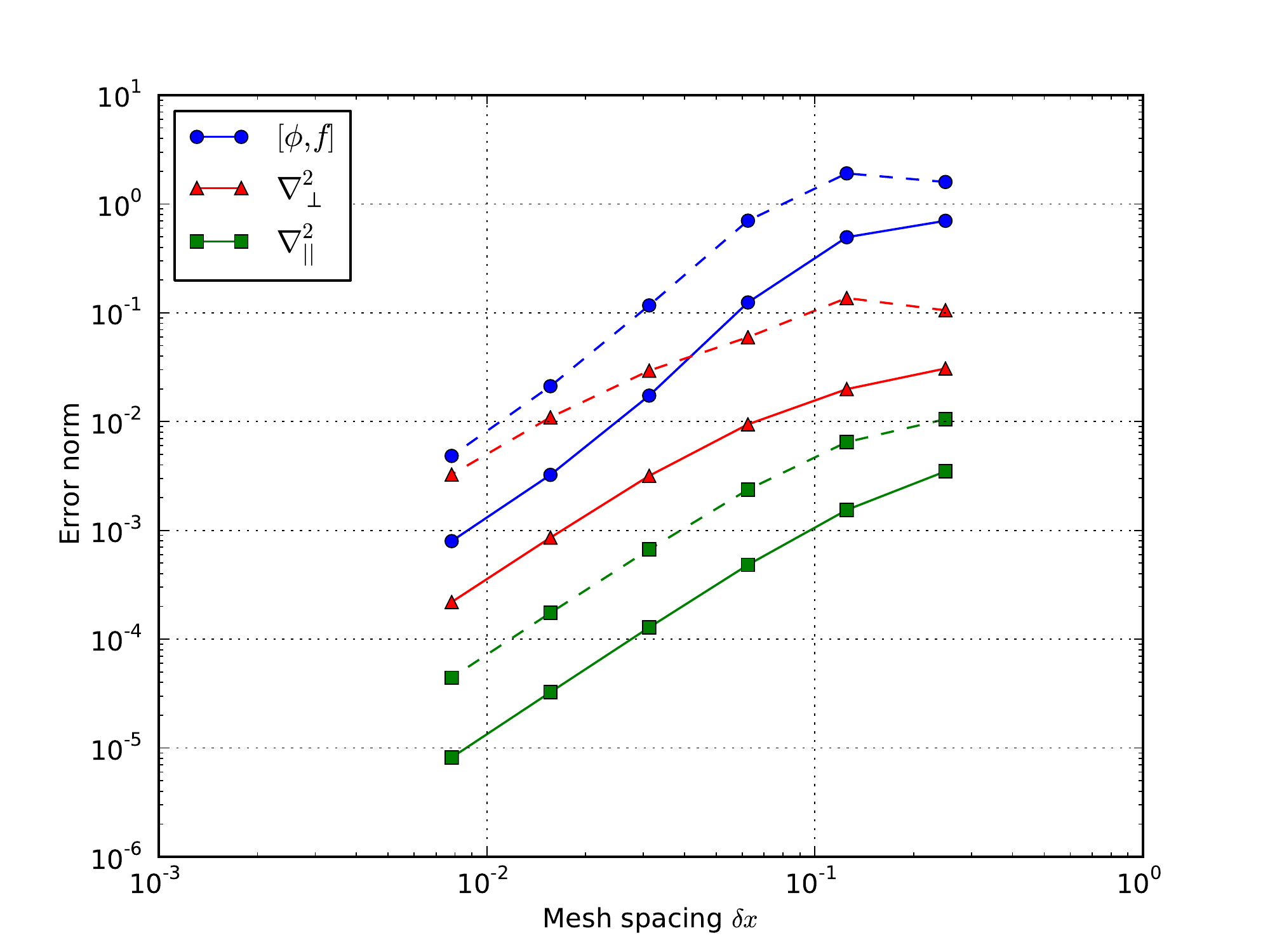}
}
\subfigure[Shifted metric method. Convergence rates for Arakawa bracket (2.07); perpendicular diffusion operator (1.98); and parallel diffusion operator (1.99)]{
  \label{fig:shifted_metric}
  \includegraphics[width=0.45\columnwidth]{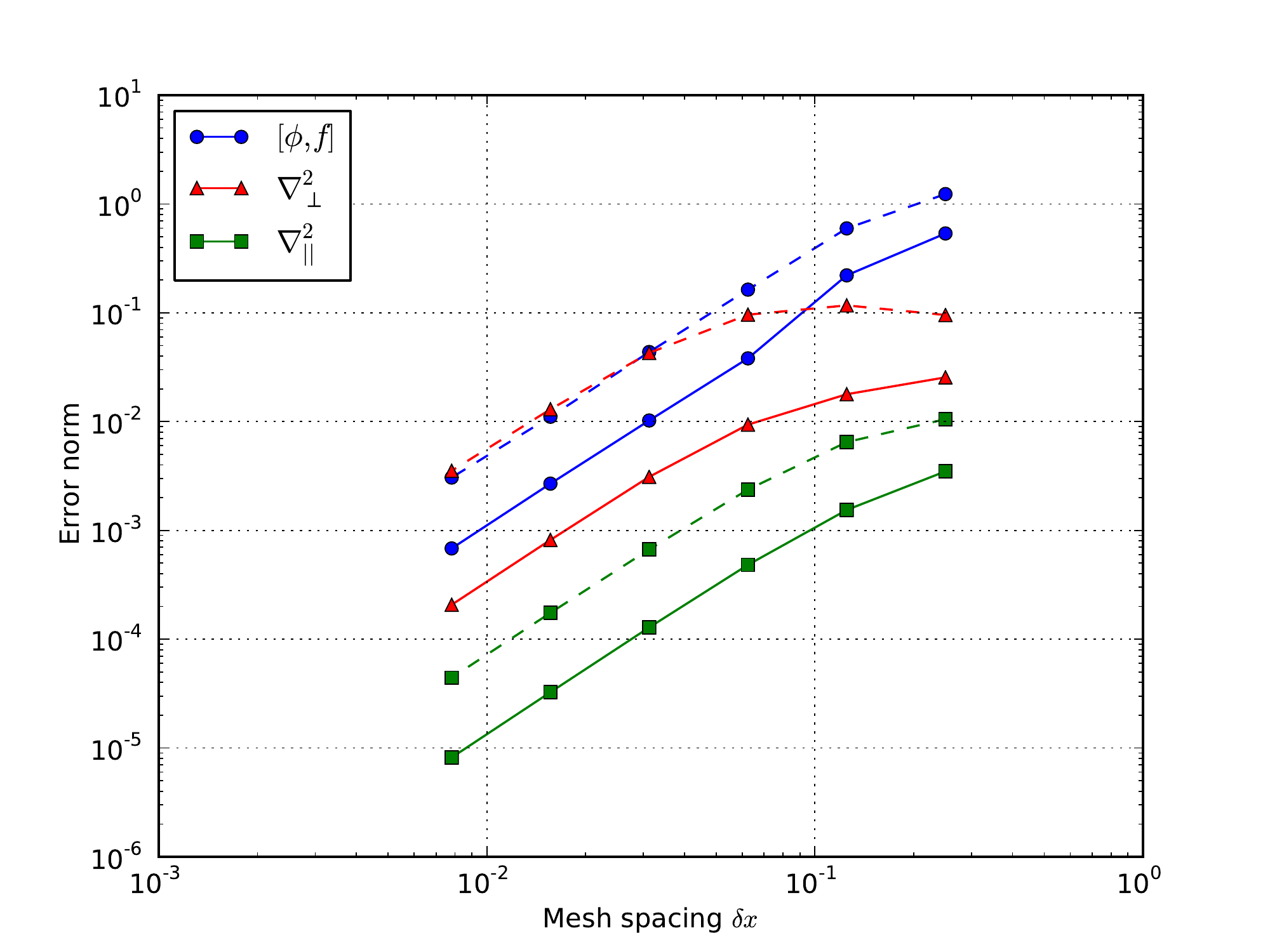}
}
\caption{Verification of operators in toroidal field-aligned coordinates. Coordinate system and input described in section~\ref{sec:coordinates}. Script:\texttt{examples/MMS/tokamak/runtest}}
\label{fig:toroidal}
\end{figure}

Results are shown in figure~\ref{fig:toroidal} for a range of resolutions from $4^3$ to $128^3$, showing convergence of the Arakawa bracket operator $\left[\phi, f\right]$, a perpendicular diffusion operator $\nabla_\perp^2$, and parallel diffusion operator $\nabla_{||}^2$. Tests in both ballooning coordinates (equations~\ref{eq:ballooning}, figure~\ref{fig:ballooning}) and shifted metric (figure~\ref{fig:shifted_metric}) show $2^{nd}$ order convergence as expected: In addition to verifying these operators in non-orthogonal curvilinear coordinates, this test exercises the twist-shift matching used to close field-lines in the core region of tokamak simulations, and the calculation of radial derivatives in the shifted metric scheme. Note that in figures~\ref{fig:ballooning} and \ref{fig:shifted_metric} the parallel diffusion operator $\nabla_{||}^2$ results are identical, as the use of shifted metrics does not affect derivatives in the parallel direction (see section~\ref{sec:coordinates}).

\begin{figure}[htbp!]
\centering
  \includegraphics[width=0.6\textwidth]{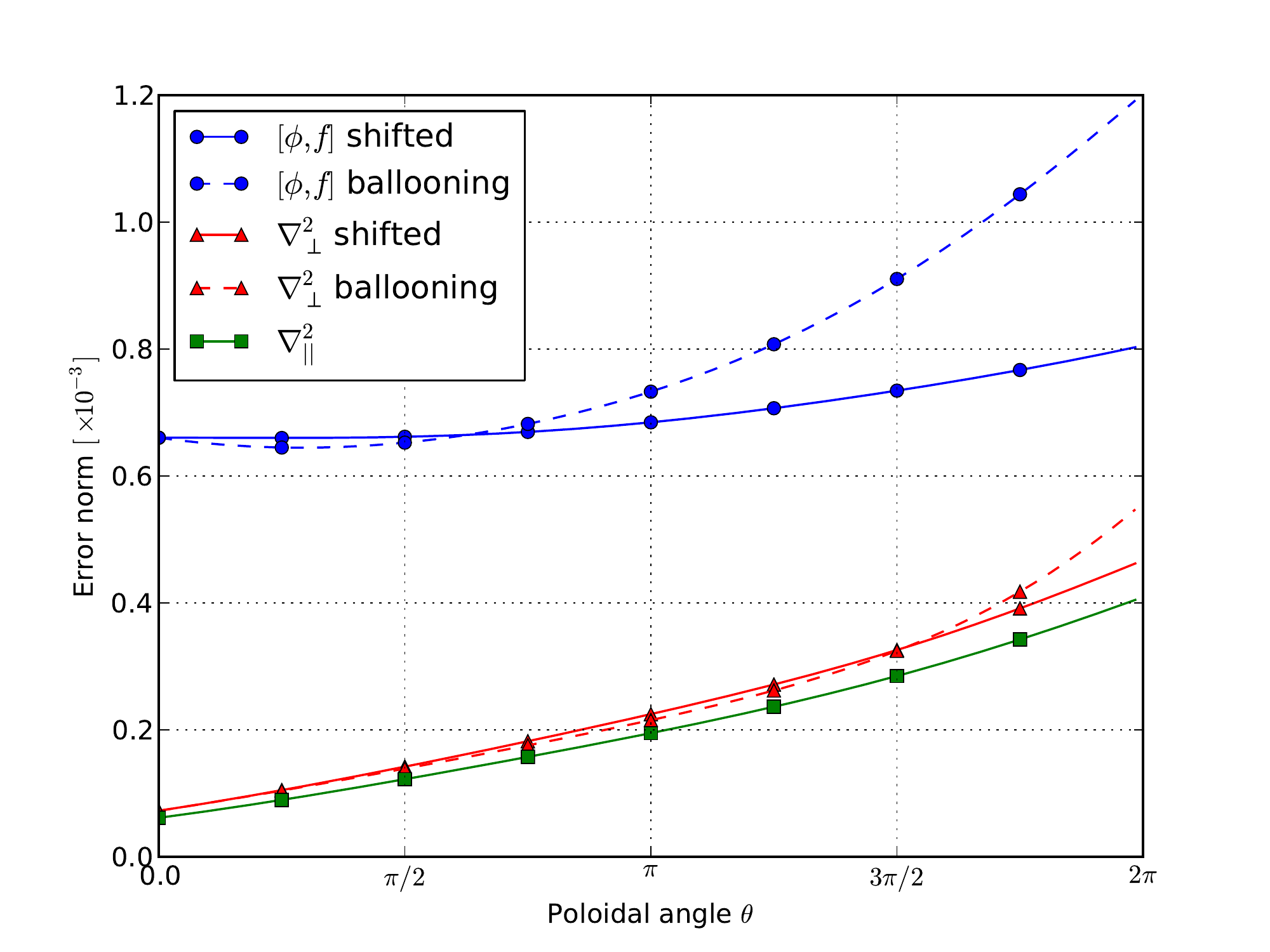}
\caption{Comparison of RMS error norm for ballooning (dashed lines) and shifted-metric scheme (solid lines), as function of poloidal angle $\theta$, for the highest resolution case in figure~\ref{fig:toroidal} ($128^3$ grid points). Integrated shear $I$ (equation~\ref{eq:sinty}) is zero at $\theta = 0$, and a maximum at $\theta = 2\pi$.}
\label{fig:norm_theta}
\end{figure}

For this test case the reference poloidal angle $\theta_0$ in equation~\ref{eq:sinty} was set to zero, so $I = 0$ at $\theta = 0$. At $\theta = 0$  the $x-z$ mesh is therefore orthogonal, and there is no difference between ballooning and shifted metric
results in figure~\ref{fig:norm_theta} at this location in $\theta$. 
Moving away from $\theta = 0$ the $x-z$ mesh becomes increasingly deformed, and differences between the ballooning and shifted-metric procedures become apparent. As expected, the error norm is largest close to $\theta = 2\pi$ where the mesh is most sheared, and the error at this point is reduced significantly by using the shifted metric procedure. 
The shifted metric method is however not always more accurate than the ballooning coordinate method, as shown for the advection operator around $\theta = \pi/4$ in figure~\ref{fig:norm_theta}, where the ballooning coordinates are more accurate: In general the accuracy of these methods will depend on the solution. It has been found in simulations of Edge Localised Modes with BOUT++~\cite{xu2010,dudson2011}, that the use of the shifted metric method improves numerical stability at the twist-shift location where the mesh deformation changes abruptly. This coordinate system is used in section~\ref{sec:3field} to verify the 3-field equations used for ELM simulations.

\subsection{Flux Coordinate Independent scheme}

To verify the interpolation and central differencing schemes implemented in BOUT++ for FCI coordinates, we simulate a wave (equation~\ref{eq:parallel_wave}) in a sheared slab. On each $x-z$ plane perpendicular to the magnetic field a Cartesian mesh is used, and the magnetic field is sheared so that the points to be interpolated (small open circles in figure~\ref{fig:fci_diagram}) span a range of locations between neighbouring grid points.

A sheared slab of size $L_y = 10$m along the magnetic field; $L_x = 0.1$m in the radial direction, and $L_z = 1$m in the binormal direction was used, with magnetic field $\left(B_x, B_y, B_z\right) = \left(0, 1, 0.05 + \left(\overline{x}-0.05\right)/10\right)$. The variation of the magnetic field-line pitch with $x$ therefore ensures that the interpolation location varies so as to test the 3$^{rd}$-order Hermite interpolation scheme. The manufactured solution used was
\begin{eqnarray}
f &=& \sin\left(\overline{y} - \overline{z}\right) + \cos\left(t\right)\sin\left(\overline{y} - 2\overline{z}\right) \\
g &=& \cos\left(\overline{y} - \overline{z}\right) + \cos\left(t\right)\sin\left(\overline{y} - 2\overline{z}\right)
\end{eqnarray}
where $\overline{y}$ and $\overline{z}$ are normalised to be between $0$ and $2\pi$ in the domain (as in all manufactured solutions presented here).

\begin{figure}[htbp!]
\centering
\includegraphics[width=0.6\textwidth]{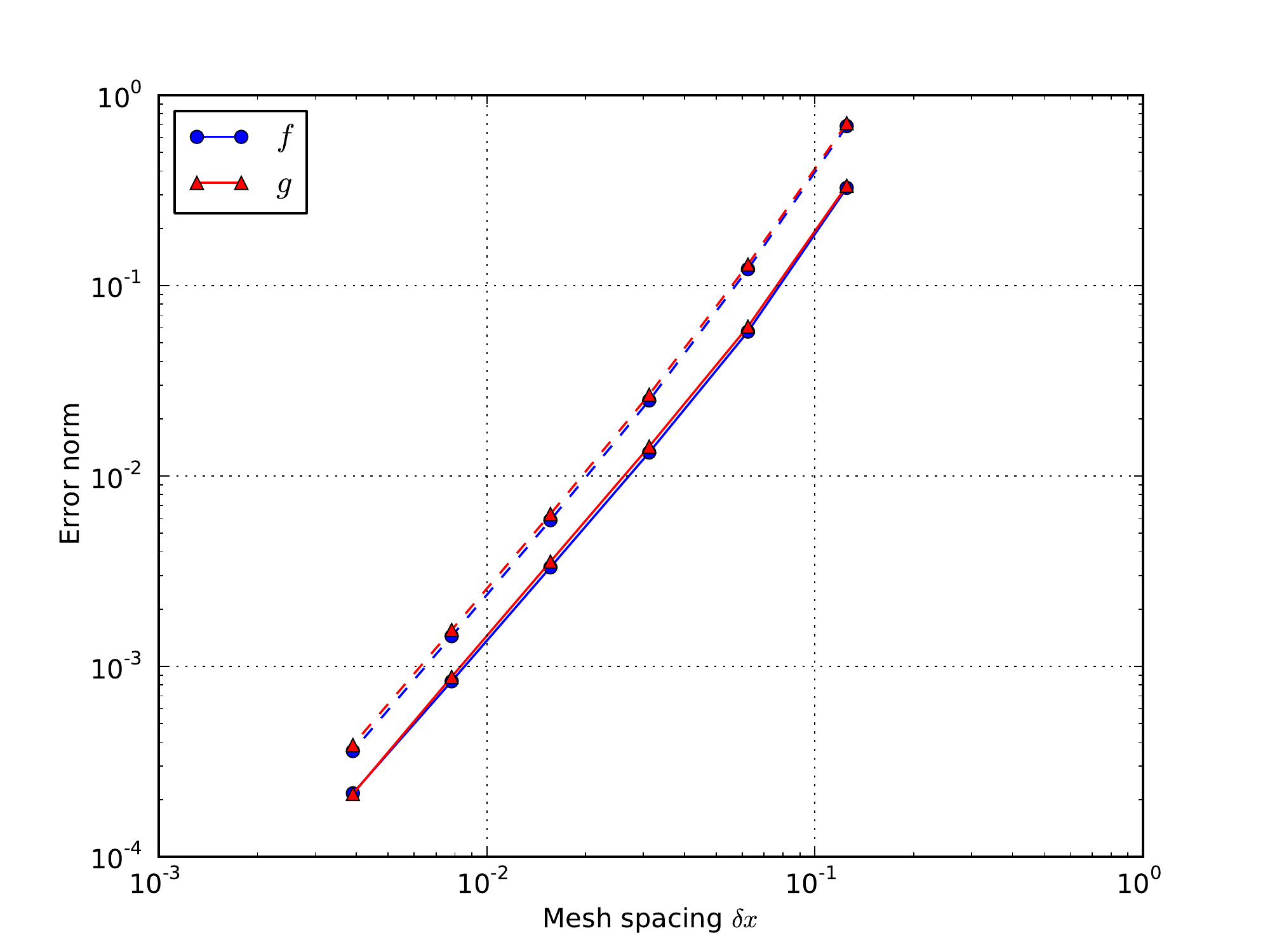}
\caption{Convergence of $f$ (1.95) and $g$ (2.04) in equation~\ref{eq:parallel_wave} solved using the FCI method in a sheared slab. Solid lines show $l^2$ (RMS) error, whilst dashed lines show $l^\infty$ (maximum) error. Script:\texttt{examples/fci-slab/runtest}}
\label{fig:fci-norm}
\end{figure}

Figure~\ref{fig:fci-norm} shows the error norm as the resolution in both parallel and perpendicular directions is varied. This shows second-order convergence, most likely limited by the accuracy of the second-order central differencing scheme used to calculate parallel derivatives. Note that in order to obtain good convergence, it was necessary to stabilise the collocated scheme, by adding a parallel diffusion term of the form $\delta x^2 \partial_{||}^2$ to each equation. This has been previously discussed in the context of MMS testing of collocated numerical schemes in~\cite{salari2000}.

\section{Models}
\label{sec:models}

After verification of individual operators, the MMS technique is now applied to the verification of entire models,
which combine operators and couple multiple fields. Here 
three models of interest are verified: the 2-field Hasegawa-Wakatani system (section~\ref{sec:hw}), a 3-field reduced MHD model which has been used extensively to simulate Edge Localised Modes (ELMs) with BOUT++ (section~\ref{sec:3field}), and
a 5-field cold-ion model for tokamak edge turbulence (section~\ref{sec:5field}).

Due to the large number of models which have been implemented in BOUT++,
we have introduced a naming scheme which can be used in future
publications to refer to a specific model. A scheme BOUT++/name/year
such as BOUT++/HW/2014 is used here.

\subsection{Hasegawa-Wakatani (BOUT++/HW/2014)}
\label{sec:hw}

 The Hasegawa-Wakatani model is a good starting place as it contains many of the elements of more complicated models, such as Poisson brackets, diffusion, and calculation of electrostatic potential from vorticity, whilst being 2-D and faster to run than 3D models at high resolutions. As such, it often forms a starting point for the construction of more complex models. The equations solved are for plasma density $n$ 
and vorticity $\omega = \mathbf{b}_0\cdot\nabla\times\mathbf{v}$ where $\mathbf{v}$ is the E$\times$B drift velocity in a constant magnetic field, and $\mathbf{b}_0$ is the unit vector in the direction of the equilibrium magnetic field:
\begin{eqnarray}
\frac{\partial n}{\partial t} &=& -\left[\phi, n\right] + \alpha\left(\phi - n\right) - \kappa\frac{\partial\phi}{\partial z} + D_n\nabla_\perp^2n \label{eq:hw}\\
\frac{\partial \omega}{\partial t} &=& -\left[\phi, \omega\right] + \alpha\left(\phi - n\right) + D_\omega\nabla_\perp^2 \omega \nonumber \\
\nabla_\perp^2\phi &=& \omega \nonumber
\end{eqnarray}
The manufactured solutions were chosen to be
\begin{eqnarray}
n &=& 0.9 + 0.9\overline{x} + 0.2\cos\left(10t\right)\sin\left(5\overline{x}^2 - 2\overline{z}\right) \\
\omega &=& 0.9 + 0.7\overline{x} + 0.2\cos\left(7t\right)\sin\left(2\overline{x}^2 - 3\overline{z}\right) \nonumber \\
\phi &=& \sin\left(\pi\overline{x}\right)\left[0.5\overline{x} - \cos\left(7t\right)\sin\left(3\overline{x}^2 - 3z\right)\right] \nonumber
\end{eqnarray}
along with parameters
\begin{equation}
\alpha = 1 \qquad \kappa = \frac{1}{2} \qquad D_n = 1 \qquad D_\omega = 1
\end{equation}
These parameters were chosen so that the magnitude of each term in equations~\ref{eq:hw} was comparable; in a realistic simulation the parameters might be different, in particular the diffusion terms $D_{n,\omega}$ would generally be smaller than is used here. This does not present a problem for verification, since the correctness of the numerical method implementation does not depend on these parameters. If the code is correct with $D_n=1$ then it will also be correct with $D_n=10^{-5}$. This does not guarantee that the method will be stable with arbitrary parameters, and in general the required resolutions and stability critera (e.g. maximum timestep) will be problem specific.

\begin{figure}[htbp!]
\centering
  \includegraphics[width=0.6\textwidth]{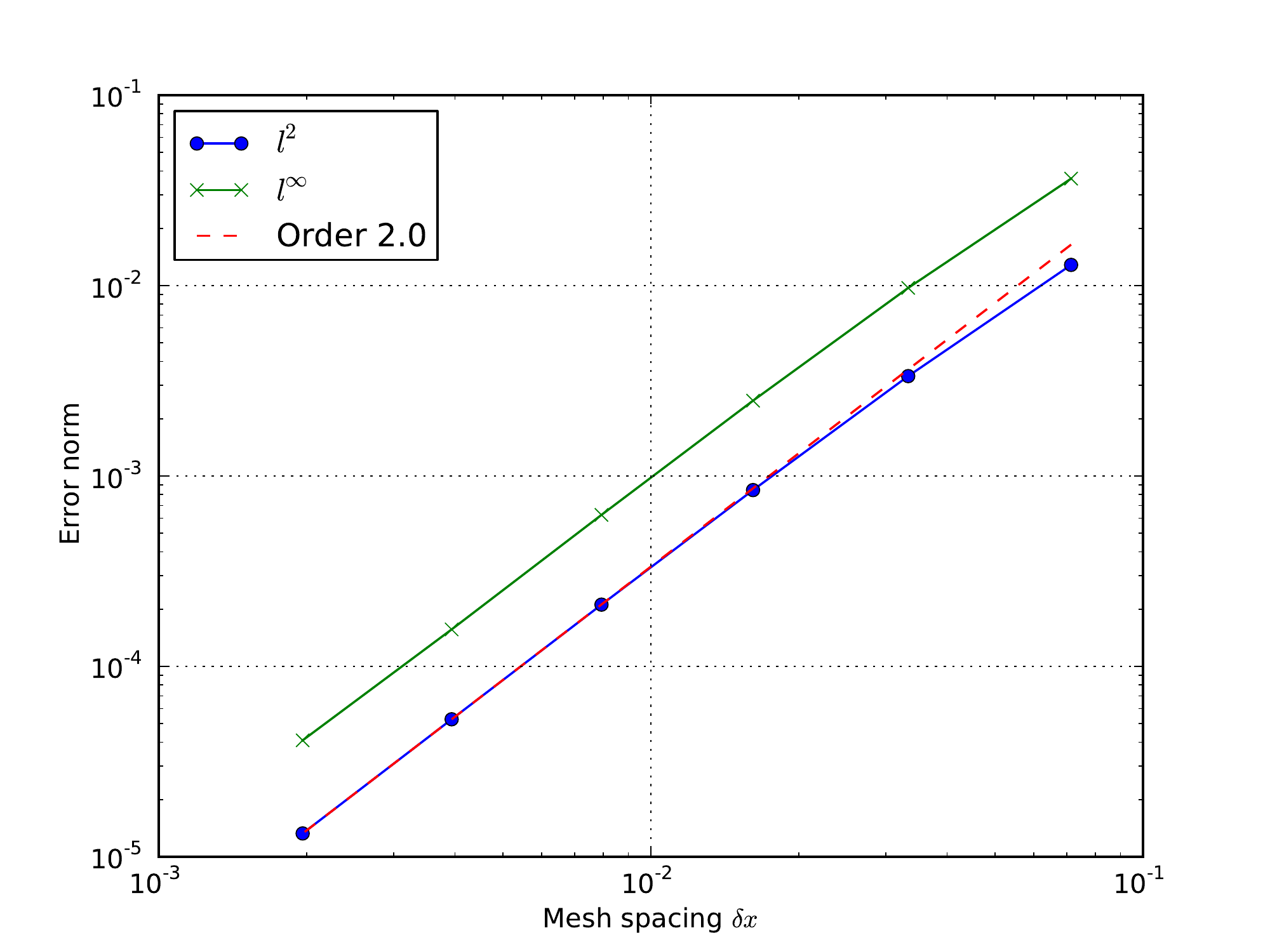}
\caption{Error norm of Hasegawa-Wakatani system ($n$ and $\phi$, equations~\ref{eq:hw}) on a Cartesian mesh, showing second-order convergence. Mesh resolutions range from $16\times 16$ to $512\times 512$. Script:\texttt{examples/MMS/hw/runtest}}
\label{fig:hw_norm}
\end{figure}

Results are shown in figure~\ref{fig:hw_norm}, calculated on a 2D unit domain, showing the
$l^2$ and $l^\infty$ norms over both $n$ and $\omega$, and a fit showing second order convergence.
This shows that the operators in equation~\ref{eq:hw} including the inversion of  potential $\phi$ from vorticity $\omega$ are correctly implemented, at least on orthogonal uniform grids. We now proceed to test these operators in toroidal field-aligned coordinate systems typical of realistic BOUT++ simulations.

\subsection{3-field reduced MHD (BOUT++/FLUID3/2014)}
\label{sec:3field}

The 3-field model used for ELM simulations~\cite{Dudson2009,xu2010,dudson2011} has been verified
in field-aligned toroidal geometry with a radially varying safety factor $q$,
using the shifted metric coordinate system described
in section~\ref{sec:coordinates}, and tested in section~\ref{sec:testcoords}.
This is in order to verify the methods implemented in BOUT++ in coordinate systems
with a non-trivial metric tensor. 

The equations evolved are for vorticity $\omega = \mathbf{b}_0\cdot\nabla\times\mathbf{v}$, pressure $p$, and the
parallel component of the magnetic vector potential $A_{||} = \mathbf{b}_0 \cdot \mathbf{A}$, where $\mathbf{b}_0 = \frac{\mathbf{B}_0}{\left|\mathbf{B_0}\right|}$; $\mathbf{B}_0$ is the unit vector along the equilibrium magnetic field $\mathbf{B}_0$, and $B_0 = \left|\mathbf{B_0}\right|$ is the magnitude of the magnetic field
\begin{eqnarray}
\rho_0 \frac{d\omega}{dt} &=& B_0^2\partial_{||}\left(\frac{\Jpar}{B_0}\right) + 2\bxk p \\
\deriv{\apar}{t} &=& -\partial_{||}\phi - \eta J_{||} \label{eq:3field_ohm}\\
\frac{dp}{dt} &=& -\frac{1}{B_0}\bvec_0\times\nabla\phi\cdot\nabla p_0 \\
\omega &=& \frac{1}{B_0}\delp\phi \\
\Jpar &=& J_{||0} - \frac{1}{\mu_0}\delp\apar
\end{eqnarray}
where the parallel derivative includes the perturbed magnetic field:
\[
\partial_{||} = \bvec_0 \cdot \nabla - \frac{1}{B_0}\bvec_0\times\nabla\apar\cdot\nabla
\]
where '$0$' subscripts denote equilibrium (starting) quantities: $\rho_0$ is the (constant) density; $\mathbf{B}_0$ the magnetic field; $\kappa_0 = \left(\mathbf{b}_0\cdot\nabla\right)\mathbf{b}$ is the field-line curvature. The electrostatic potential $\phi$ is calculated from the vorticity by inverting a perpendicular Laplacian (with Dirichlet boundary conditions here), and the parallel current $\Jpar = \mathbf{b}_0 \cdot\mathbf{J}$ is calculated from the vector potential. The convective derivative is defined as
\begin{equation}
\frac{d}{dt} = \deriv{}{t} + \frac{1}{B_0}\bvec_0\times\nabla\phi\cdot\nabla
\end{equation}
Background (equilibrium) profiles are chosen
to mimic realistic cases, with a pedestal-like pressure profile $P_0$, and a
parallel current profile $J_0$ which peaks on the outboard and inboard midplanes:
\begin{equation}
P_0/\overline{P} = 2 + \cos\left(\pi \overline{x}\right) \qquad J_0/\overline{J} = 1 - \overline{x} + \sin^2(\pi\overline{x})\cos\left(\theta\right)
\end{equation}
where $\overline{x}$ is the normalised radial coordinate, which lies between $0$ and $1$, and $\theta$ is the poloidal angle, which lies between $0$ and $2\pi$. 
Normalisation parameters are
\begin{eqnarray}
n_e &=& 10^{19}~\textrm{m}^{-3} \qquad T_e = 3~\textrm{eV} \qquad \overline{L} = 1\textrm{m} \qquad \overline{B} = 1\textrm{T}\\
\overline{P} &=& 2en_eT_e = 9.6~\textrm{Pa} \qquad \overline{J} = \overline{B} /\left(\mu_0 \overline{L} \right) \nonumber
\end{eqnarray}
The manufactured solutions used were:
\begin{eqnarray}
\phi &=& \left[\sin\left(\overline{z} - \overline{x} + t\right) + 10^{-3}\cos\left(\overline{y} - \overline{z}\right)\right
]\sin\left(2\pi \overline{x}\right) \\
\psi &=& 10^{-4}\cos\left(4\overline{x}^2 + \overline{z} - \overline{y}\right) \\
U  &=& 2\sin\left(2t\right)\cos\left(\overline{x} - \overline{z} + 4\overline{y}\right) \\
P  &=& 1 + \frac{1}{2}\cos\left(t\right)\cos\left(3\overline{x}^2 - 2\overline{z}\right) + 5\times 10^{-3}\sin\left(\overline{y}-\overline{z}\right)\sin\left(t\right)
\end{eqnarray} 

A Lundquist number of $S = 10$ was used to set the resistivity $\eta$. This is so that the resistive term in Ohm's law (equation~\ref{eq:3field_ohm}) becomes comparable to the other terms, and $S$ is much smaller (higher $\eta$) than would be the case in a realistic tokamak simulation, for which $S = 10^8$ would be more typical.

\begin{figure}[htbp!]
\centering
  \includegraphics[width=0.6\textwidth]{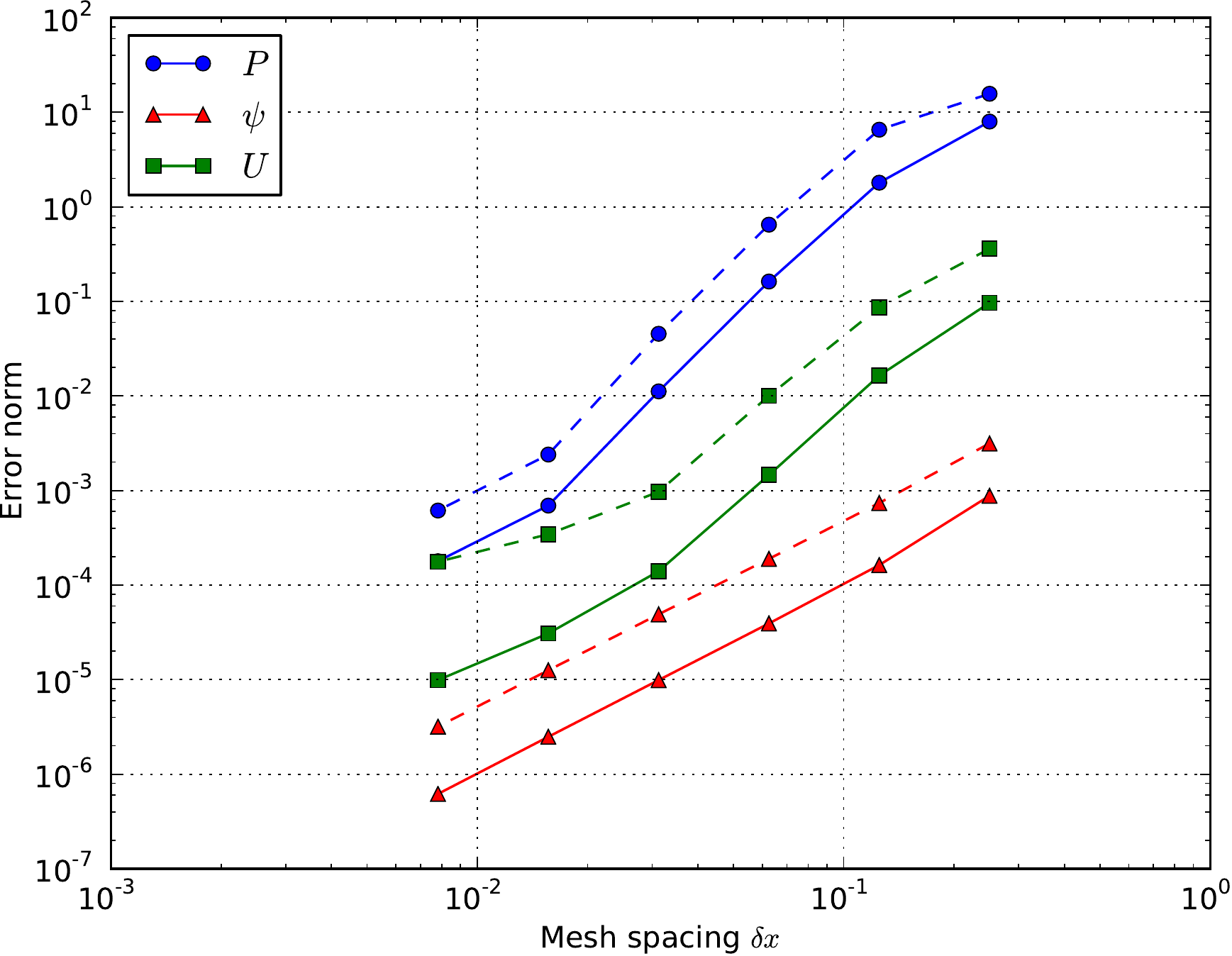}
\caption{Error norms for 3-field set of equations. Solid lines show the $l^2$ (RMS) error norms, whilst dashed lines are the $l^\infty$ (maximum) error. Convergence orders for pressure $p$ is 1.95; vorticity $\omega$ is 1.64; and vector potential $\apar$ is 2.01. Resolutions range from $4^3$ to $128^3$. Sctript:\texttt{examples/MMS/elm-pb/runtest}}
\label{fig:elm-pb-norm}
\end{figure}

Results are shown in figure~\ref{fig:elm-pb-norm}, with the $l^2$ and $l^\infty$ norms shown for each evolving variable $\left(P, \psi, U\right)$.
The slow convergence at large mesh spacing (small resolution) is due to the solutions being under-resolved: the smallest grids have only $4$ grid points in each dimension, insufficient to resolve the manufactured solution. At high resolution the pressure and
electromagnetic potential fields converge at 2$^{nd}$ order as expected, but the vorticity $\omega$ converges at a rate between first and second order. The maximum ($l^\infty$) error in vorticity converges at close to $1^{st}$ order at high resolution, indicating
that the source of this slow convergence is an order $1$ error on a sub-set of the domain, so that when averaged over the domain the RMS ($l^2$) error converges at a faster rate than the maximum error. The location of the error maximum at high resolution is at the radial boundary, but the reason for this is not yet clear despite extensive investigation. Here we conclude that although the model does converge, it does not converge at the expected rate, and further investigation is needed.

\subsection{BOUT++/FLUID5/2014}
\label{sec:5field}

Finally, the set of equations implemented in the Global Braginskii Solver (GBS) code~\cite{ricci2012} have been implemented in BOUT++ and verified in a simplified form using the Method of Manufactured Solutions. In this current work electromagnetic effects and ion viscosity terms were neglected. The equations are for plasma density $n$, electron temperature $T_e$, vorticity $\omega$, Ohm's law, and parallel ion velocity $V_{||i}$:
\begin{eqnarray}
  \deriv{n}{t} &=& -\frac{R}{\rho_{s0}}\frac{1}{B}\left[\phi, n\right] + \frac{2n}{B}\left[C\left(T_e\right) + \frac{T}{n}C\left(n\right) - C\left(\phi\right)\right] \nonumber \\
    &&- n\left(\bvec\cdot\nabla\right)V_{||e} - V_{||e}\left(\bvec\cdot\nabla\right)n + D\left(n\right) + S \label{eq:gbs_n}\\
\deriv{T_e}{t} &=& -\frac{R}{\rho_{s0}}\frac{1}{B}\left[\phi, T_e\right] - V_{||e}\bdotGrad T_e \nonumber \\
    &&+ \frac{4}{3}\frac{T_e}{B}\left[\frac{7}{2}C\left(T_e\right) + \frac{T_e}{n}C\left(n\right) - C\left(\phi\right)\right] \nonumber \\
    &&+ \frac{2T_e}{3}\Bigg[0.71\bdotGrad V_{||i} - 1.71\bdotGrad V_{||e} \nonumber \\
      &&  + 0.71 \frac{\left(V_{||i} - V_{||e}\right)}{n}\bdotGrad n\Bigg] \nonumber \\
    && + D_{T_e}\left(T_e\right) + D_{T_e}^{||}\left(T_e\right) + S_T \\
  \deriv{\omega}{t} &=& -\frac{R}{\rho_{s0}}\frac{1}{B}\left[\phi, \omega\right] - V_{||i}\left(\bvec\cdot\nabla\right)\omega \nonumber \\
  &&+ B^2\left[\bdotGrad\left(V_{||i} - V_{||e}\right) + \frac{\left(V_{||i} - V_{||e}\right)}{n}\bdotGrad n\right]\nonumber  \\
  &&+ 2B\left[ C\left(T_e\right) + \frac{T_e}{n}C\left(n\right)\right] + D_\omega\left(\omega\right) \\
  \deriv{V_{||e}}{t} &=& -\frac{R}{\rho_{s0}}\frac{1}{B}\left[\phi, V_{||e}\right] - V_{||e}\bdotGrad V_{||e} \nonumber \\
    &-& \frac{m_i}{m_e}\nu\left(V_{||e} - V_{||i}\right) + \frac{m_i}{m_e}\bdotGrad\phi \nonumber \\
  &-&\frac{m_iT_e}{nm_e}\bdotGrad n - 1.71\frac{m_i}{m_e}\bdotGrad T_e + D_{V_{||e}}\left(V_{||e}\right) \\
  \deriv{V_{||i}}{t} &=& -\frac{R}{\rho_{s0}}\frac{1}{B}\left[\phi, V_{||i}\right] - V_{||i}\bdotGrad V_{||i} \nonumber \\
  &&- \bdotGrad T_e + \frac{T_e}{n}\bdotGrad n + D_{V_{||i}}\left(V_{||i}\right) \label{eq:gbs_vi}
\end{eqnarray}
where
\begin{eqnarray}
\rho_{s0} &=& \frac{C_{s0}}{\Omega_{ci}} \qquad C_{s0} = \sqrt{\frac{e\overline{T}_e}{m_i}} \qquad \Omega_{ci} = \frac{e\overline{B}}{m_i}
\end{eqnarray}
with vorticity and the curvature operator defined as
\begin{equation}
\omega = \nabla_\perp^2 \phi \qquad C\left(A\right) = \frac{B}{2}\left(\nabla\times\frac{\bvec}{B}\right)\cdot\nabla A
\end{equation}
Here the dissipation operators $D\left(\cdot\right)$ were hyper-diffusion terms in the plane perpendicular to the magnetic field of the form:
\begin{equation}
D\left(f\right) = -\delta x^4 \frac{\partial^4 f}{\partial x^4} - \delta z^4 \frac{\partial^4 f}{\partial z^4}
\end{equation}

In order to test all terms in this set of equations, the parameters of the
simulation should be chosen so that the magnitude of each term is of a similar
order of magnitude. If this is not done, then the error in the result will be
dominated by a small number of operators, and mistakes in the implementation of
small terms may not become apparent until very high (possibly impractical)
resolution is reached. In order to handle the large number of terms in equations~\ref{eq:gbs_n}-\ref{eq:gbs_vi}, the magnitude of each term was estimated using SymPy by replacing trigonometric functions $\sin\left(\cdot\right)$ and $\cos\left(\cdot\right)$ by their maximum value ($1$), and the coordinates $\left(\overline{x}, \theta, \zeta\right)$ by their maximum values $\left(1, 2\pi, 2\pi\right)$. This allowed parameters to be quickly adjusted to find useful regimes. The resulting manufactured solutions are:
\begin{eqnarray}
n &=& 0.9 + 0.9\overline{x} + 0.5\cos\left(t\right)\sin\left(5\overline{x}^2 - z\right) + 0.01*\sin\left(y - z\right) \nonumber \\
T_e &=& 1 + 0.5\cos\left(t\right)\cos\left(3\overline{x}^2 - 2z\right) + 0.005\sin\left(y-z\right)\sin\left(t\right) \nonumber \\
\omega &=& 2\sin\left(2t\right)\cos\left(x - z + 4y\right) \nonumber \\
V_e &=& \cos\left(1.5t\right)\left[2\sin\left(\left(\overline{x}-0.5\right)^2 + z\right) +  0.05\cos\left(3x^2 + y - z\right)\right] \\
V_i &=& -0.01\cos\left(7t\right)\cos\left(3\overline{x}^2 + 2y - 2z\right) \nonumber \\
\phi &=& \left[\sin\left(z - \overline{x} + t\right) + 0.001\cos\left(y - z\right)\right]\sin\left(2\pi\overline{x}\right) \nonumber \\
\end{eqnarray}
Parameters used were:
\begin{eqnarray}
\overline{T}_e &=& 3\textrm{eV} \qquad \overline{n}_e = 10^{19}\textrm{m}^{-3} \qquad \overline{B} = 0.1\textrm{T} \qquad m_i = 0.1m_p
\end{eqnarray}
where $m_p$ is the mass of the proton. Light ions were used in order to reduce the difference in timescales between electrons and ion dynamics. Note that the manufactured solutions and parameters are not required to be realistic, provided that they do not violate any constraints such as positivity of density and temperature, as discussed in section~\ref{sec:testing}.

\begin{figure}[htbp!]
\centering
  \includegraphics[width=0.6\textwidth]{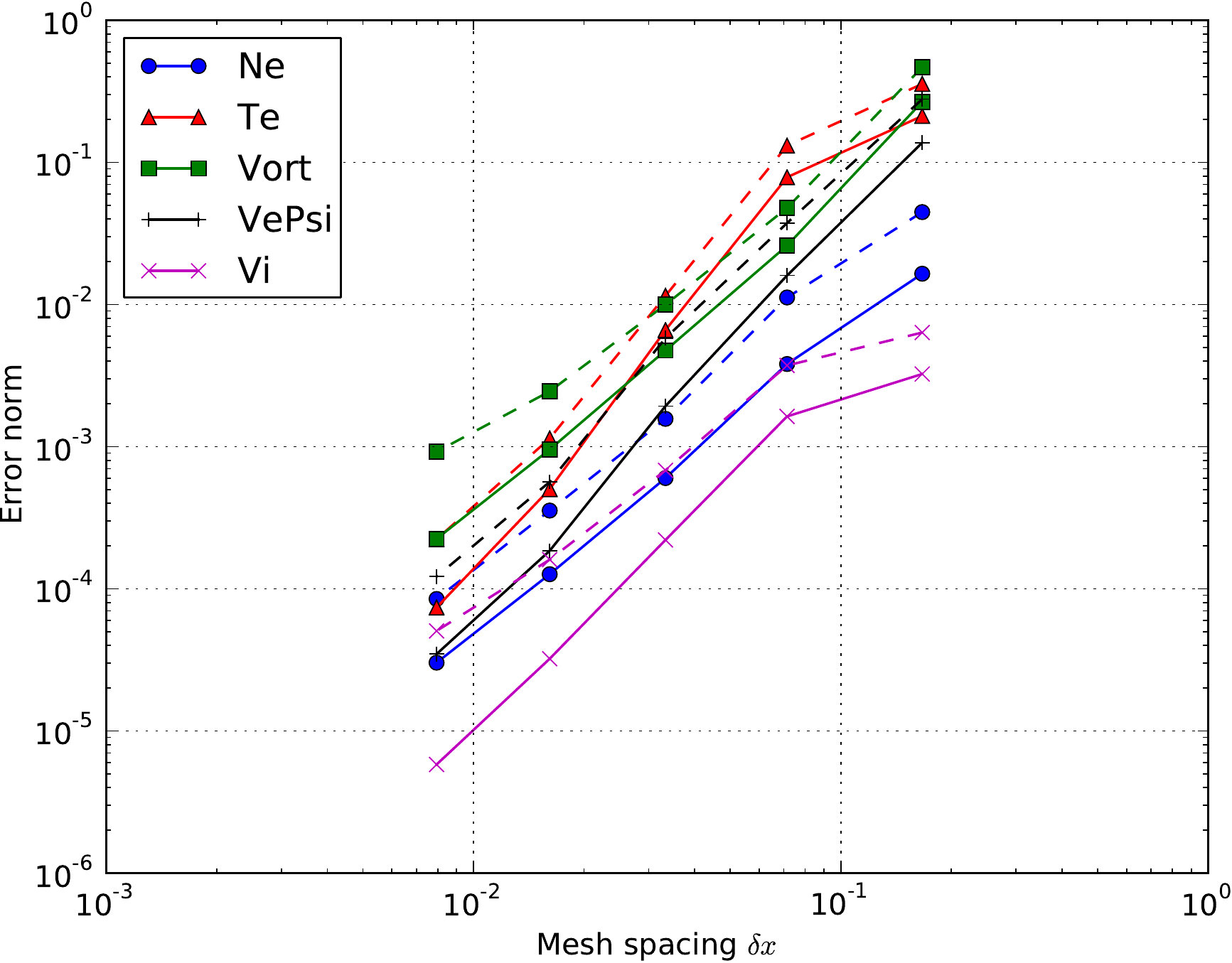}
\caption{Error norms for 5-field set of equations. Solid lines show the $l^2$ (RMS) error norms, whilst dashed lines are the $l^\infty$ (maximum) error. Convergence orders for density $N_e$ is 2.02; electron temperatue $T_e$ is 2.70; Vorticity $\omega$  is 2.04; Electron parallel velocity $V_e$ is 2.36; and ion velocity $V_i$ is 2.42. Resolutions range from $8^3$ to $128^3$. Script:\texttt{examples/MMS/GBS/runtest-slab3d}}
\label{fig:norm_slab3d}
\end{figure}

Simulations were performed in a 3D slab geometry, with resulting error
norms shown in figure~\ref{fig:norm_slab3d}. In this geometry the
curvature polarisation vector $\nabla\times\frac{\bvec}{B}$ is set to 
a constant in the $z$ (binormal) direction.
All fields show convergence at the expected rate, approximately 2$^{nd}$-order in
mesh spacing $\delta x$. This demonstrates that complex models can be verified using the method of manufactured solutions in BOUT++.

\section{Conclusions and discussion}

The Method of Manufactured solutions has been used to rigorously test
numerical methods implemented in BOUT++, both independently as unit tests,
and in combination as simulation models. Convergence to the correct
solution at an asymptotic $2^{nd}$ order has been demonstrated for large sub-sets of the BOUT++ framework: Though
higher order methods ($3^{rd}$-order WENO and $4^{th}$-order central differencing) are implemented in BOUT++, the overall convergence rate is limited to $2^{nd}$ order by the boundary conditions.

Mechanisms have been implemented into BOUT++, which simplify and partly 
automate the process of verifying the correctness of a numerical
implementation, requiring minimal modifications to the code
between production simulations and verification runs.
This will facilitate the routine use of the MMS as an increasing variety of models are implemented in BOUT++.
Since code verification is an ongoing process, particularly for an actively developed scientific code such as BOUT++,
the methods and tests detailed here are now used as part of a test suite which is run routinely and automatically (using Travis-CI) to test every change made to BOUT++. 

It is important to note the limitations of the present work, which will be
the subject of further development. Whilst curvilinear coordinates in tokamak geometry with varying safety factor have been verified, no tests have yet been performed in X-point geometry. The Flux Coordinate Independent (FCI) scheme has been implemented in BOUT++, but only tested in sheared slab geometry. Investigation of methods for simulations of X-point geometry, including FCI, and verification with MMS will be the subject of future work. 

\section*{Acknowledgements}

This work has been carried out within the framework of the EUROfusion Consortium and has received funding from the Euratom research and training programme 2014-2018 under grant agreement No 633053. The views and opinions expressed herein do not necessarily reflect those of the European Commission. The authors gratefully acknowledge the support of the UK Engineering and Physical Sciences Research Council (EPSRC) under grant EP/K006940/1, and Archer computing resources under Plasma HEC consortium grant EP/L000237/1.

\section*{References}
\bibliography{bout-mms}

\begin{thebibliography}{10}

\bibitem{Dudson2009}
B~D Dudson et~al.
\newblock {\em Comp. Phys. Comm.}, 180:1467--1480, 2009.

\bibitem{dudson2014}
B~D Dudson et~al.
\newblock {\em J. Plasma Phys.}, 81(01):365810104, 2015.
\newblock doi:10.1017/S0022377814000816.

\bibitem{roache1998}
P~J Roache.
\newblock {\em Verification and Validation in Computational Science and
  Engineering}.
\newblock Hermosa Publishers, Albuquerque NM, 1998.

\bibitem{oberkampf2010}
W~L Oberkampf and C~J Roy.
\newblock {\em Verification and Validation in Scientific Computing}.
\newblock Cambridge University Press, New York, {NY}, {USA}, 2010.

\bibitem{salari2000}
K~Salari and P~Knupp.
\newblock Code verification by the method of manufactured solutions.
\newblock Technical Report SAND2000-1444, Sandia National Laboratories, 2000.

\bibitem{umansky-2008-tests}
M~V Umansky, R~H Cohen, L~L LoDestro, and X~Q Xu.
\newblock {\em Contrib. Plasma Phys.}, 48(1-3):27--31, 2008.
\newblock http://dx.doi.org/10.1002/ctpp.200810004.

\bibitem{roy2004}
C~J Roy, C~C Nelson, T~M Smith, and C~C Ober.
\newblock {\em Int. J. Num. Methods in Fluids}, 44(6):599--620, 2004.

\bibitem{kalupin2008}
D~Kalupin et~al.
\newblock In {\em Europhysics Conference Abstracts (Proc. of the 35th {EPS}
  Conference on Plasma Physics, Hersonissos, Crete, 2008)}, volume 32D, pages
  P--5.027, 2008.

\bibitem{chang2009}
C~S Chang et~al.
\newblock {\em J. Phys.: Conf. Ser.}, 180:012057, 2009.

\bibitem{riva2014}
F~Riva et~al.
\newblock {\em Physics of Plasmas}, 21:062301, 2014.

\bibitem{michoski2014}
C~Michoski, D~Meyerson, T~Isaac, and F~Waelbroeck.
\newblock Discontinuous galerkin methods for plasma physics in the scrape-off
  layer of tokamaks.
\newblock {\em J. Comput. Phys.}, 274:898--919, 2014.

\bibitem{sympy}
{SymPy Development Team}.
\newblock {\em SymPy: Python library for symbolic mathematics}, 2014.

\bibitem{leveque2007}
R~LeVeque.
\newblock {\em Finite Difference Methods for Ordinary and Partial Differential
  Equations}.
\newblock SIAM, 2007.

\bibitem{haeseler-1}
W~D Haeseler.
\newblock {\em Flux Coordinates and Magnetic Field Structure}.
\newblock Springer, 1991.

\bibitem{xu-2008}
X~Q Xu, M~V Umansky, B~Dudson, and P~B Snyder.
\newblock Boundary plasma turbulence simulations for tokamaks.
\newblock {\em Comm. in Comput. Phys.}, 4(5):pp. 949--979, November 2008.

\bibitem{dimits-1993}
A~M Dimits.
\newblock {\em Phys. Rev. E}, 48(5):4070--4079, Nov 1993.

\bibitem{scott01}
B~Scott.
\newblock {\em Physics of Plasmas}, 8(2):447, 2001.

\bibitem{hariri2013}
F~Hariri and M~Ottaviani.
\newblock {\em Comp. Phys. Comm.}, 184(11):2419--2429, 2013.

\bibitem{stegmeir2014}
A~Stegmeir, D~Coster, O~Maj, and K~Lackner.
\newblock {\em Contrib. Plasma Phys.}, 54:549--554, 2014.

\bibitem{iserles2009}
Areih Iserles.
\newblock {\em A First Course in the Numerical Analysis of Differential
  Equations}.
\newblock Cambridge University Press, 2009.
\newblock {ISBN}: 978-0-521-73490-5.

\bibitem{karniadakis1991}
G~E Karniadakis, M~Israeli, and S~A Orszag.
\newblock {\em J. Comput. Phys.}, 97:414, 1991.

\bibitem{scott2005-arxiv}
B~D Scott.
\newblock {GEM} - an energy conserving electromagnetic gyrofluid model.
\newblock {\em arXiv:physics}, page 0501124, 2005.

\bibitem{gottlieb2001}
S~Gottlieb, C-W Shu, and E~Tadmor.
\newblock {\em {SIAM Review}}, 43(1):89--112, 2001.

\bibitem{hindmarsh2005}
A~C Hindmarsh et~al.
\newblock {SUNDIALS}: Suite of nonlinear and differential/algebraic equation
  solvers.
\newblock {\em {ACM} Transactions on Mathematical Software}, 31(3):363--396,
  2005.

\bibitem{efficient}
Satish Balay, William~D. Gropp, Lois~Curfman McInnes, and Barry~F. Smith.
\newblock In E.~Arge, A.~M. Bruaset, and H.~P. Langtangen, editors, {\em Modern
  Software Tools in Scientific Computing}, pages 163--202. Birkhauser Press,
  1997.

\bibitem{petsc-user-ref}
S~Balay et~al.
\newblock Technical Report ANL-95/11 - Revision 3.1, Argonne National
  Laboratory, 2010.

\bibitem{arakawa1960}
Arakawa. A.
\newblock {\em J. Comput. Phys.}, 1:119--143, 1960.

\bibitem{jiang-1996}
Guang-Shan Jiang and Chi-Wang Shu.
\newblock {\em J. Comput. Phys.}, 126:202--228, 1996.

\bibitem{jiang-1997}
Guang-Shan Jiang and Danping Peng.
\newblock {\em {SIAM} J. Sci. Comp.}, 21(6):2126--2143, 2000.

\bibitem{xu2010}
X~Q Xu et~al.
\newblock {\em Phys. Rev. Lett.}, 105:175005, 2010.

\bibitem{dudson2011}
B~D Dudson et~al.
\newblock {\em Plasma Phys. Control. Fusion}, 53:054005, 2011.
\newblock doi: 10.1088/0741-3335/53/5/054005.

\bibitem{ricci2012}
P~Ricci et~al.
\newblock {\em Plasma Phys. Control. Fusion}, 54:124047, 2012.

\end{thebibliography}
\bibliographystyle{unsrt}

\end{document}